\documentclass[lettersize,journal]{IEEEtran}
\usepackage{amsmath,amsfonts}
\usepackage{algorithmic}
\usepackage{algorithm}
\usepackage{array}
\usepackage[caption=false,font=normalsize,labelfont=sf,textfont=sf]{subfig}
\usepackage{textcomp}
\usepackage{stfloats}
\usepackage{url}
\usepackage{verbatim}
\usepackage{graphicx}
\usepackage{booktabs}
\usepackage{tabularx}
\usepackage{multirow}
\usepackage{amssymb}
\usepackage{bm}
\usepackage[noadjust]{cite} 
\usepackage[colorlinks,linkcolor=blue,anchorcolor=blue,citecolor=blue,urlcolor=black]{hyperref}
\usepackage{xcolor}
\usepackage{soul}
\soulregister\cite7
\soulregister\ref7

\newsavebox{\localizationtablebox}
\newsavebox{\featuretablebox}
\newsavebox{\relatedsurveytablebox}
\newcolumntype{Y}{>{\raggedright\arraybackslash}X}
\newcommand{\lowstar}{\ensuremath{\bigstar}}
\newcommand{\medstar}{\ensuremath{\bigstar\bigstar\bigstar}}
\newcommand{\highstar}{\ensuremath{\bigstar\bigstar\bigstar\bigstar\bigstar}}
\newcommand{\lowtri}{\ensuremath{\blacktriangle}}
\newcommand{\medtri}{\ensuremath{\blacktriangle\blacktriangle\blacktriangle}}
\newcommand{\hightri}{\ensuremath{\blacktriangle\blacktriangle\blacktriangle\blacktriangle\blacktriangle}}
\newcommand{\ratingcirc}{\raisebox{0.08ex}{\scalebox{1.25}{\ensuremath{\bullet}}}}
\newcommand{\lowcirc}{\ratingcirc}
\newcommand{\medcirc}{\ratingcirc\ratingcirc\ratingcirc}
\newcommand{\highcirc}{\ratingcirc\ratingcirc\ratingcirc\ratingcirc\ratingcirc}
\newcommand{\lowsq}{\ensuremath{\blacklozenge}}
\newcommand{\medsq}{\ensuremath{\blacklozenge\blacklozenge\blacklozenge}}
\newcommand{\highsq}{\ensuremath{\blacklozenge\blacklozenge\blacklozenge\blacklozenge\blacklozenge}}
\newcommand{\targetdata}[2]{\makebox[\linewidth][c]{\makebox[2.8em][c]{#1}\,/\,\makebox[4.4em][l]{#2}}}

\begin{document}

\title{Learning-Driven Channel Representation for Wireless Localization: From Channel Observations to Location Inference}

\author{Hongyu Xie, Chenglong Li, Xinming Huang, Emmeric Tanghe,~\IEEEmembership{Member,~IEEE}, David Plets,~\IEEEmembership{Senior Member,~IEEE}, Wout Joseph,~\IEEEmembership{Senior Member,~IEEE}, Shaojie Ni, and Xiaojun Yuan,~\IEEEmembership{Fellow,~IEEE}

\thanks{H. Xie, C. Li, X. Huang and S. Ni are with the College of Electronic Science and Technology, National University of Defense Technology, 410073 Changsha, China, and also with the National Key Laboratory for Positioning, Navigation and Timing Technology, 410073 Changsha, China (e-mail: \{xiehongyu, chenglong.li\}@nudt.edu.cn, nishaojie123@126.com, hxm\_kd@163.com).}

\thanks{E. Tanghe, D. Plets and W. Joseph are with the WAVES, Department of Information Technology, Ghent University, 9052 Ghent, Belgium (e-mail: \{emmeric.tanghe, david.plets and wout.joseph\}@ugent.be).}

\thanks{Xiaojun Yuan is with the National Key Laboratory of Wireless Communications, University of Electronic Science and Technology of China, 611731 Chengdu, China (e-mail: xjyuan@uestc.edu.cn).}
}



\maketitle

\begin{abstract}
Wireless observations capture radio signal responses formed through interactions with propagation environments and spatial geometry. In integrated sensing and communication, such observations have become an important basis for high-accuracy localization beyond conventional channel estimation. Learning-driven methods learn implicit relations between channel propagation and spatial position, enabling location inference under complex channel conditions. However, the useful information is tightly coupled with spatial layout, environmental dynamics, and system configurations. This coupling obscures the inference process and weakens performance consistency across scenarios. In this paper, we review learning-driven high-accuracy localization techniques with channel representations as the organizing view. The survey covers typical channel observation forms and analyzes their physical meanings. We also review channel feature extraction and representation learning methods, and summarize methods according to the acquisition, organization, adaptation, and reuse of channel representations. Typical methods are compared in terms of accuracy, applicable conditions, data requirements, and generalization. We model the localization process as a unified ``wireless observation--channel representation--location inference'' framework and use information bottleneck theory to highlight that the quality and usability of channel representations are critical to exploiting propagation information and play a decisive role in localization performance. Finally, we summarize the key challenges in moving from experimental studies to real deployment and present our perspectives on these issues. 
\end{abstract}

\begin{IEEEkeywords}
Channel representation, channel feature, learning-driven, wireless localization, channel observation.
\end{IEEEkeywords}

\section{Introduction}
\IEEEPARstart{H}{igh} precision positioning is evolving from terminal-oriented location services into a fundamental capability of future wireless networks \cite{behravan2022positioning}. Rapid progress in B5G/6G has given rise to many emerging applications, such as the Internet-of-Things (IoT), extended reality, autonomous vehicles, and intelligent robots. In these applications, location information must support real-time decisions and spatial interaction with high accuracy and reliability. Moreover, localization systems need to accommodate more diverse deployment environments, exhibiting superior adaptability in terms of coverage, terminal density, power constraints, and infrastructure costs \cite{3gpp22261}. These requirements make reliable and scalable localization an unavoidable challenge in future wireless networks.

Integrated sensing and communication (ISAC) creates new opportunities for localization. By integrating sensing functions into wireless communication networks, ISAC allows communication, sensing, and localization to reuse spectrum and network infrastructure, while exploiting their mutual gains \cite{liu2022integrated}, \cite{3gpp22137}. Specifically, communication procedures continuously generate channel observations, which provide a measurement basis for ubiquitous sensing services \cite{dai2026tutorial}. By analyzing signal patterns and propagation characteristics, wireless networks can support sensing and localization with existing infrastructure. In turn, location and environmental awareness can assist beamforming, channel prediction, and resource management, reducing real-time channel acquisition and reference signal overhead while improving link robustness \cite{liu2022integrated}, \cite{koivisto2017high}. Therefore, ISAC enables wireless networks to provide high-quality and robust communication services while supporting real-time, accurate sensing within a unified system architecture.

Learning-driven wireless positioning changes how wireless observations are used. Previous studies have shown that machine learning and deep networks can learn location-related features from complex channel observations, enabling effective fingerprinting localization \cite{wang2016deepfi}, non-line-of-sight (NLoS) error mitigation \cite{bregar2018improving}, and end-to-end localization \cite{hsieh2019deep}. More importantly, learning-driven methods are not restricted to handcrafted channel statistics or geometric parameters. They can automatically extract more structured and discriminative channel features from high-dimensional wireless measurements. This allows propagation information related to spatial location to be used more effectively by localization models. Recently, self-supervised learning \cite{salihu2024self} and pretrained models \cite{ott2024radio} have started to learn general wireless representations from unlabeled channel data. Such representations can support downstream localization tasks with limited labeled samples.

\begin{table*}[t]
\caption{Comparison of surveys related to wireless localization.}
\label{tab:related_surveys}
\centering
\scriptsize
\renewcommand{\arraystretch}{1.2}
\setlength{\tabcolsep}{2.5pt}
\begin{lrbox}{\relatedsurveytablebox}
\begin{tabularx}{\textwidth}{c|l|Y|c|c|c|c}
\hline
\multicolumn{1}{c|}{
    \multirow{3}{*}{\textbf{Ref.}}
} &
\multicolumn{1}{c|}{
    \multirow{3}{*}{\textbf{Scope}}
} &
\multicolumn{1}{c|}{
    \multirow{3}{*}{
        \shortstack[c]{
            \textbf{Organizing Perspective}
        }
    }
} &
\multicolumn{1}{c|}{
    \multirow{3}{*}{
        \shortstack[c]{
            \textbf{Representation}\\
            \textbf{Learning}
        }
    }
} &
\multicolumn{3}{c}{
    \textbf{Effects of Representations on Localization}
} \\

\cline{5-7}

& & & &
\textbf{Localization} &
\textbf{Data} &
\textbf{Generalization} \\

& & & &
\textbf{Accuracy} &
\textbf{Requirements} &
\textbf{Capability} \\
\hline\hline
\cite{zafari2019survey} & Indoor localization & Systems and technologies & $\sim$ & $\checkmark$ & $\sim$ & $\times$ \\
\cite{sallouha2024ground} & Ground--air--space localization & Anchor/target types and geometry & $\times$ & $\checkmark$ & $\times$ & $\times$ \\
\cite{vaezi2022cellular} & Cellular, wide-area, and non-terrestrial IoT & IoT technologies and performance metrics & $\times$ & $\sim$ & $\times$ & $\times$ \\
\cite{stock2024survey} & Opportunistic LEO-PNT & Receiver processing and error sources & $\times$ & $\checkmark$ & $\sim$ & $\times$ \\
\cite{italiano2024tutorial} & 5G localization & Standards, signals, and positioning & $\sim$ & $\checkmark$ & $\times$ & $\times$ \\
\cite{trevlakis2023localization} & 6G localization & Enabling technologies and methods & $\sim$ & $\sim$ & $\checkmark$ & $\checkmark$ \\
\cite{khalajmehrabadi2017modern} & WLAN fingerprint localization & Algorithms and deployment challenges & $\sim$ & $\checkmark$ & $\checkmark$ & $\checkmark$ \\
\cite{zhu2020indoor} & Intelligent fingerprint localization & Fingerprint types and learning algorithms & $\sim$ & $\checkmark$ & $\checkmark$ & $\checkmark$ \\
\cite{fischer2025systematic} & Angle-based localization & Angle estimation and system design & $\sim$ & $\checkmark$ & $\sim$ & $\times$ \\
\cite{guo2019survey} & Fusion-based indoor localization & Information sources, algorithms, and fusion weights & $\sim$ & $\sim$ & $\sim$ & $\sim$ \\
\cite{shit2019ubiquitous} & Device-free localization & Measurements, methods, and systems & $\sim$ & $\checkmark$ & $\checkmark$ & $\checkmark$ \\
\cite{burghal2020comprehensive} & Machine learning for wireless localization & Measurements, features, and learning pipelines & $\sim$ & $\checkmark$ & $\checkmark$ & $\checkmark$ \\
\cite{pan2025ai} & AI-driven wireless positioning & Standards, parameter estimation, and direct positioning & $\sim$ & $\checkmark$ & $\checkmark$ & $\checkmark$ \\
\cite{wang2026tutorial} & Learning-based radio map construction & Data, learning paradigms, and physics & $\sim$ & $\sim$ & $\sim$ & $\times$ \\
\cite{umer2025reconfigurable} & RIS-assisted localization & System models, scenarios, and algorithms & $\sim$ & $\checkmark$ & $\sim$ & $\sim$ \\
\cite{zeng2024tutorial} & Channel knowledge map & Map construction, channel inference, and applications & $\sim$ & $\sim$ & $\sim$ & $\sim$ \\
\cite{ren2026channel} & Channel knowledge map construction & Interpolation, deep learning, and neural fields & $\sim$ & $\times$ & $\times$ & $\times$ \\
\cite{huachaca2026wireless} & Wireless foundation models & Pretraining, adaptation, and generalization & $\sim$ & $\sim$ & $\checkmark$ & $\checkmark$ \\
\cite{cheraghinia2026wireless} & Physical-layer foundation models & Architectures, learning, applications, and deployment & $\sim$ & $\sim$ & $\checkmark$ & $\checkmark$ \\
\hline
\textbf{Ours} & Channel representation-based localization & \parbox[c]{\linewidth}{\raggedright\strut Channel representation, unified framework, information bottleneck, generalization\strut} & $\checkmark$ & $\checkmark$ & $\checkmark$ & $\checkmark$ \\
\hline
\end{tabularx}
\end{lrbox}
\usebox{\relatedsurveytablebox}
\par\vspace{2pt}
\parbox{\wd\relatedsurveytablebox}{\raggedright\scriptsize
\textbf{\textit{Note:}} $\checkmark$ denotes systematic substantive discussion. $\sim$ denotes partial substantive coverage. $\times$ denotes no substantive discussion.\par
\noindent Representation learning covers channel feature extraction and representation learning.\par
\noindent Localization accuracy concerns representation effects on localization accuracy.\par
\noindent Data requirements cover data and labels for construction, training, and adaptation.\par
\noindent Generalization capability covers representation reuse and adaptation across tasks or data distributions.}
\end{table*}

Despite the promising performance and potential of learning-driven methods in wireless positioning, the process from wireless observations to location inference should not be simply interpreted as an end-to-end mapping. In fact, channel observations contain location-related information, such as propagation distance, multipath structure (coupled with surrounding layout), and spatial geometry, but this information is coupled with unstable factors caused by scene layout, system configuration, and dynamic disturbances. Therefore, localization accuracy and generalization depend not only on the final inference module, but also on whether the model can extract and preserve effective location-related information from wireless observations. To this end, increasing research efforts have focused on enhancing location-related representations through the design of model architectures, loss functions, and training procedures.

Several recent surveys have reviewed wireless positioning and its supporting techniques from three main perspectives: application scenarios, observation types, and methodologies. From the application perspective, existing reviews discuss positioning under specific environments or network deployments, such as indoor localization \cite{zafari2019survey}, ground-air-space network localization \cite{sallouha2024ground}, cellular, wide-area, and non-terrestrial IoT \cite{vaezi2022cellular}, opportunistic PNT with signals from LEO communication satellites \cite{stock2024survey}, and 5G positioning and 6G localization \cite{italiano2024tutorial}, \cite{trevlakis2023localization}. Other reviews are organized around the measurements used for localization, including WiFi fingerprinting \cite{khalajmehrabadi2017modern}, \cite{zhu2020indoor}, angle-based localization \cite{fischer2025systematic}, multi-source fusion localization \cite{guo2019survey}, and device-free localization \cite{shit2019ubiquitous}. Method-oriented surveys further summarize machine learning and AI-driven localization \cite{burghal2020comprehensive}, \cite{pan2025ai}, radio map construction \cite{wang2026tutorial}, RIS-assisted localization \cite{umer2025reconfigurable}, channel knowledge maps \cite{zeng2024tutorial}, \cite{ren2026channel}, and wireless foundation models \cite{huachaca2026wireless}, \cite{cheraghinia2026wireless}. Table~\ref{tab:related_surveys} compares these surveys with this work in terms of scope, organizing perspective, and coverage of representation learning and its effects on localization. While these works offer valuable insights from diverse perspectives, they predominantly categorize existing studies by application scenarios, measurement modalities, or algorithmic architectures. Consequently, the underlying mapping mechanism bridging wireless observations and location inference remains insufficiently explored. To the best of our knowledge, this work is the first to foreground this mapping paradigm as the central focus of learning-driven wireless localization, systematically investigating its pivotal role in the inference pipeline.

To this end, this paper revisits the process from wireless observations to localization. We define channel representations as features extracted or learned from wireless observations that characterize channel propagation structure and location-related information. The process of obtaining such representations from wireless observations is referred to as channel representation learning. Location inference refers to using these representations for geometric parameter estimation, position regression, or fingerprint-based localization. Based on this view, learning-driven localization is modeled as a unified ``wireless observation--channel representation--location inference'' framework. Different from existing surveys, this paper reviews high-precision wireless positioning from the perspective of channel representation. We cover typical channel observation forms, channel feature extraction and representation learning methods, and channel representation-based localization methods, followed by a discussion of the key challenges. The main contributions of this paper are as follows.

\begin{enumerate}
\item We establish a unified analytical framework for learning-driven wireless positioning. The framework consists of wireless observation, channel representation, and location inference. It revisits the localization process by examining how channel representations are formed and how they support location inference, providing a more interpretable basis for learning-driven localization.
\item We review typical channel observation forms, channel feature extraction methods, and representation learning strategies in wireless positioning, highlighting the location-related information and expression characteristics of different observations and representations.
\item We systematize representation-based localization according to the acquisition mechanisms and structural organization of channel representations, offering a comparative analysis of their distinct characteristics in terms of localization accuracy, applicability, and generalizability.
\item We outline the critical challenges hindering the transition of representation-based localization from laboratory setups to real-world deployments, providing valuable perspectives for upcoming research efforts.
\end{enumerate}

\begin{figure*}[t]
\centerline{\includegraphics[width=0.95\textwidth]{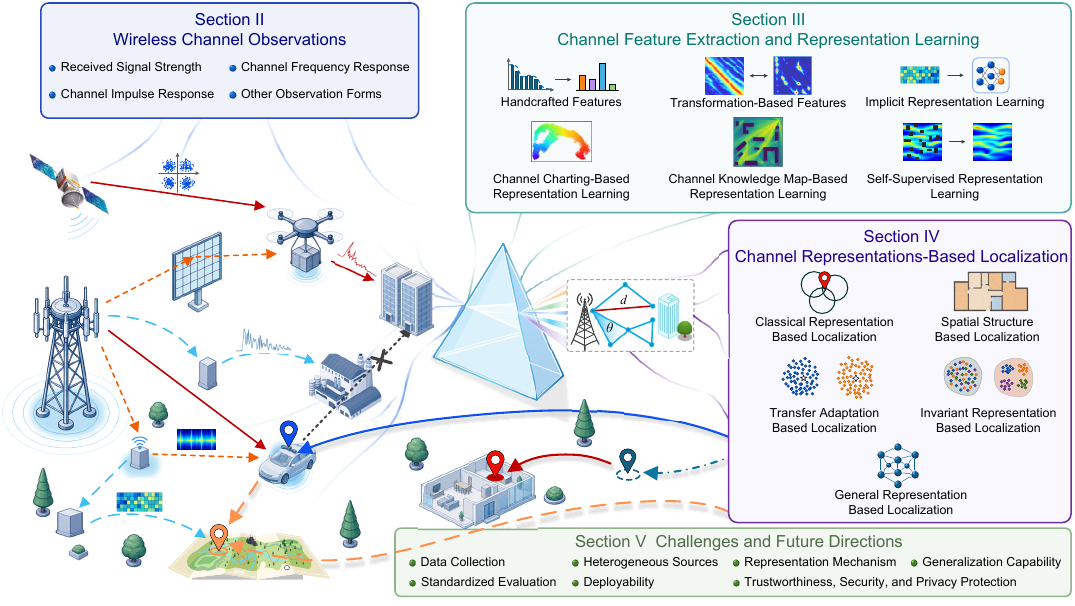}}
\caption{Survey organization and section overview.}
\vspace{-2mm}
\label{fig:survey_organization}
\end{figure*}

Fig.~\ref{fig:survey_organization} outlines the scope and organization of this survey. Section~\ref{sec:channel_observations} introduces typical channel observation forms in wireless positioning. Section~\ref{sec:feature_representation_learning} reviews channel feature extraction and representation learning methods, including handcrafted features, transform-domain features, implicit representations, channel charting, channel knowledge maps, and self-supervised learning. Section~\ref{sec:representation_based_localization} summarizes channel representation-based localization methods, analyzes their localization performance and generalization, and concludes with a unified localization framework and an information bottleneck analysis. Section~\ref{sec:challenges} discusses critical bottlenecks and deployment challenges. Section~\ref{sec:conclusion} concludes this survey.

\section{Wireless Channel Observations}
\label{sec:channel_observations}

\begin{figure}[t]
\centerline{\includegraphics[width=\linewidth]{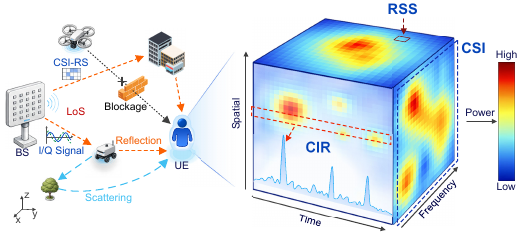}}
\caption{Typical channel observations for channel representation.}
\vspace{-2mm}
\label{fig:channel_observations}
\end{figure}

The physical channel is a key carrier that links spatial location, propagation environment, and channel observations. As wireless signals propagate through space, they experience not only path loss, but also complex interactions with the environment via blockage, reflection, and scattering. Consequently, the channel response is closely related to the spatial environment. Wireless positioning essentially infers the spatial geometric information of a target by observing the propagation characteristics between transceivers \cite{gentner2016multipath}, \cite{witrisal2016high}. Channel observations provide a source of localization information, but the different observation forms illustrated in Fig.~\ref{fig:channel_observations} vary significantly in dimensionality and information granularity, yielding distinct profiles of channel propagation characteristics and disparate degrees of location sensitivity. Accordingly, this section introduces typical channel observation forms, analyzes the physical sources of localization-related geometric information and their main influencing factors.

\subsection{Received Signal Strength}

Received signal strength (RSS) reflects the power of the signal received at the receiver. Under free-space propagation, the received power can be described by the Friis transmission equation \cite{sarkar2003survey}. However, practical wireless channels are affected by complex environmental factors such as obstruction, reflection, and scattering, rendering the free-space assumption inapplicable. Therefore, the log-distance path loss model is commonly used in engineering practice to describe the average received signal power relative to the transmit power \cite{andersen1995propagation}:

\begin{equation}
P_r(d)=P_r(d_0)-10n \log_{10}(\frac{d}{d_0})+X_{\sigma}, d>d_0
\end{equation}
where $P_r(d)$ and $P_r(d_0)$ denote the received power at distance $d$ and reference distance $d_0$, respectively, $n$ is the path loss exponent, and $X_{\sigma}$ is the random perturbation due to shadowing.

RSS is widely used in localization because it reflects signal attenuation and is easily obtained. However, RSS also has inherent limitations in localization tasks \cite{yang2013rssi}, \cite{wang2016phasefi}. First, as a low-dimensional and coarse-grained measurement reflecting received power, RSS cannot describe the multipath structure and rich spatiotemporal information contained in wireless signals. Second, RSS is sensitive to non-line-of-sight (NLoS) propagation, multipath fading, and time-varying channels, and its statistical properties can shift over time and across scenarios. These issues prevent RSS from supporting accurate environmental and channel representations and confine RSS-based localization mainly to meter-level accuracy \cite{yang2013rssi}.

\subsection{Channel Impulse Response}

The channel impulse response (CIR) characterizes the time-domain response of a wireless channel. In multipath environments, the received signal can be modeled as the superposition of multiple propagation components, and the CIR can be expressed as \cite{molisch2004ieee}:
\begin{equation}
h(\tau)=\sum_{l=1}^{L}\alpha_l\delta(\tau-\tau_l),
\end{equation}
where $L$ is the number of propagation paths, $\alpha_l$ and $\tau_l$ denote the complex amplitude and delay of the $l$th path component, respectively, and $\delta(\cdot)$ is the Dirac delta function. 

Compared to RSS, the CIR provides a finer-grained characterization of propagation paths and environmental geometry by recording the energy attenuation and delay distribution of the propagation process. Its location-related information can be summarized in two aspects. Path delays provide geometric constraints on propagation distance \cite{yang2023uwb}, while the delay structure of multipath components reflects channel conditions and spatial layout \cite{angarano2021robust}, \cite{leitinger2015evaluation}. Consequently, CIR can constrain the user position through time-of-arrival (ToA) measurements \cite{kuhn2010adaptive}, \cite{coene2023noise}, or support location inference through the association between overall waveform features and spatial locations \cite{stahlke2022transfer}, \cite{fontaine2023ultra}. The environmental information contained in CIR can further support sensing and localization \cite{sun2022channel}, \cite{coene2024location}.

Although CIR provides richer channel information, it still faces practical limitations \cite{kim2020review}, \cite{makela2024ubiquitous}. First, the representation capability of CIR is constrained by the physical bandwidth of the wireless system, as temporal resolution is inversely proportional to bandwidth. Under such constraints, dense multipath components may overlap in the delay domain, weakening the ability of CIR to resolve the multipath structure. Second, under NLoS conditions, the line-of-sight (LoS) path can be severely attenuated or even completely blocked, making first-path detection difficult and introducing positive bias into ToA-based ranging. Third, CIR-based localization systems typically require large bandwidth and high-precision time synchronization, which increases system complexity and deployment cost.

\subsection{Channel Frequency Response}

The channel frequency response (CFR) characterizes the frequency-domain response of a wireless channel. It is the frequency-domain equivalent of the CIR, and the two are linked by the Fourier transform (FT) \cite{stefanoni2025survey}. Consider a multiple-input multiple-output (MIMO) orthogonal frequency division multiplexing (OFDM) system where the base station (BS) is equipped with a uniform planar array consisting of $N_{\mathrm{ant}}$ antennas, and the user equipment (UE) is equipped with a single omnidirectional antenna. The system has $N_{\mathrm{sub}}$ subcarriers, and the CFR of the $k$th subcarrier can be expressed as \cite{heath2016overview}:
\begin{equation}
\mathbf{h}_k
=
\sum_{l=1}^{L}
\alpha_l\mathbf{a}(\theta_l,\phi_l)
e^{-\mathrm{j}2\pi f_k\tau_l},
\end{equation}
where $L$ is the number of propagation paths, $\alpha_l$ and $\tau_l$ are the complex gain and propagation delay of the $l$th multipath signal, respectively. $f_k$ is the frequency of the $k$th subcarrier. $\mathbf{a}(\theta_l,\phi_l)$ represents the array response, where $\theta_l$ and $\phi_l$ denote the azimuth and elevation angles of arrival (AoA), respectively. The array channel frequency responses across subcarriers can be stacked column-wise to form the channel state information (CSI) matrix: $\mathbf{H}=[\mathbf{h}_1,\mathbf{h}_2,\dots,\mathbf{h}_{N_{\mathrm{sub}}}]\in\mathbb{C}^{N_{\mathrm{ant}}\times N_{\mathrm{sub}}}$.

CSI preserves richer amplitude and phase information. Its amplitude reflects the response strength resulting from the coherent superposition of multipath components across subcarriers and antenna elements, while its phase records the phase rotations across subcarriers caused by propagation delays and the spatial phase differences across antenna elements caused by arrival directions \cite{yang2013rssi}, \cite{ruan2022hi}. Consequently, CSI not only characterizes the frequency and spatial selectivity of the channel, but also contains location-related information associated with propagation delays, arrival directions, and multipath structure \cite{kotaru2015spotfi}, \cite{feng2018wireless}. The propagation paths and scatterer distributions generally vary with spatial locations, and these discrepancies translate into distinct, discriminative response patterns across subcarriers and antennas \cite{ruan2022hi}, \cite{kim2021unsupervised}, \cite{liu2023machine}. Applying the FT to CSI further provides an explicit representation of propagation paths in the angular-delay domain.

However, using CSI for localization remains challenging \cite{kim2020review}, \cite{feng2018wireless}. First, the high dimensionality of CSI makes conventional model-based super-resolution methods computationally demanding, while its performance is constrained by idealized channel and array model assumptions. Moreover, synchronization errors and hardware impairments introduce various frequency-dependent phase errors and phase offsets across antenna elements, so the raw phase cannot accurately describe position-dependent information and is not directly usable. Furthermore, data-driven methods typically require large amounts of accurately labeled data, resulting in high collection costs. When the environmental layout, human activity, or system configuration changes, the existing fingerprint database or trained model may become outdated or unreliable, requiring new data collection, database updates, or model retraining.

\subsection{Other Observation Forms}

Communication systems also generate physical-layer observations during protocol interaction and data transmission, including in-phase/quadrature (I/Q) signals, reference signals, and pilots. These signals are not necessarily designed for localization, but are produced for synchronization, channel estimation, and link maintenance. However, their waveforms, phases, and time-frequency structures are also shaped by propagation attenuation, multipath scattering, and spatial geometry. They therefore carry propagation information and can support location inference with limited additional overhead.

I/Q signals are obtained by down-converting and sampling the received signals and retain waveform and time-frequency variations before channel estimation. RF-GPT \cite{zou2026rf} converts I/Q waveforms into time-frequency spectrograms and uses a vision encoder and a language model to learn time-frequency structures and protocol-related features in radio-frequency (RF) signals, showing the potential of raw communication waveforms as a learnable wireless observation modality. Because I/Q signals are close to raw received signals, their observations contain not only propagation-induced waveform variations, but also non-propagation factors related to hardware implementation, synchronization offsets, signal formats, and payload information. Models may mistake these differences related to the system or service for location-related features.

Reference and pilot signals in communication are another important source of observations. With known signal structures and relatively controllable resource positions, these signals provide more stable propagation observations than ordinary traffic data and reduce the overhead of dedicated localization measurements. In 5G new radio (5G NR), the positioning reference signal (PRS), channel state information reference signal (CSI-RS), sounding reference signal (SRS), and synchronization signal block (SSB) can provide location-related information \cite{sallouha2024ground}, \cite{pan2025ai}. However, their observation quality is constrained by reference-signal bandwidth, transmission periodicity, resource density, and network synchronization, so they cannot always reliably support high-accuracy localization.

\subsection{Summary}
Overall, channel responses characterize wireless propagation across the delay, frequency, and spatial dimensions, providing rich geometric and multipath information for location inference. However, this information cannot always be directly resolved and extracted from raw observations for localization. Communication-related observations further expand the information sources available for localization and can reduce additional measurement overhead by reusing existing communication resources, but they also contain non-propagation factors related to protocols, hardware, and services. Therefore, feature extraction or representation learning is needed to extract stable and discriminative channel information from different observation forms and support subsequent tasks.

\section{Channel Feature Extraction and Representation Learning}
\label{sec:feature_representation_learning}

\subsection{Why Channel Representations Matters?}

Channel representation refers to the feature representation of the channel propagation structure, extracted or learned from wireless observations, which encapsulates location-dependent information. Unlike parameterized methods that rely on exact signal models, representation learning does not require pre-establishing a complete observation model. Furthermore, it does not strictly depend on far-field/near-field assumptions, priors on the number of paths, array manifolds, channel sparsity, or signal-to-noise ratio (SNR) conditions. It focuses on organizing and refining location-related propagation structures in channel measurements through learning, turning complex observations into clear location-related representations.

The basic structure of a wireless channel is mainly determined by physical factors such as the propagation scenario, propagation medium, and user location, and electromagnetic propagation gives rise to different propagation-path characteristics. Therefore, location-related information is retained in signal measurements as power attenuation, delay distribution, phase variation, angular response, or time-frequency structure. However, raw channel measurements are essentially direct observations of the wireless propagation environment and cannot be directly equated with stable and generalizable geometric representations. The location-related information in these raw measurements is tightly coupled with environmental dynamics \cite{jiao2025robust}, hardware imperfections \cite{tubail2022effect}, and system configuration differences \cite{zafari2019survey}. These measurements also differ in dimensionality, information granularity, and physical interpretation \cite{zafari2019survey}, \cite{pan2025ai}. Conventional data-driven localization methods based solely on channel measurements often overlook the physical nature of wireless channels. As a result, many localization methods work only under scenario-specific system configurations and task settings \cite{gong2023deep}, and their performance may degrade sharply when the physical domain changes or shifts. From channel measurements to location inference, raw observations need to be transformed through a channel representation process into representations of channel information that support location discrimination and inference.

Channel feature extraction aims to transform raw channel measurements into effective representations. Existing methods have progressed from handcrafted features to implicit mappings learned by data-driven models, and more recently to physically guided designs that integrate propagation priors with model architectures. This evolution has increased both the depth of representation learning and the ability to disentangle useful location-related information from irrelevant factors. This section reviews the literature on channel feature extraction and representation learning.

\subsection{Handcrafted Features}

Handcrafted features are a basic form of representation in channel feature extraction.
Wireless signals experience different propagation processes, and their channel measurements show distinct energy distributions, multipath structures, and waveform patterns. The received signal or channel response also exhibits different amplitude, phase, and time-frequency statistics under different channel states. Based on these differences, channel feature parameters can be computed and selected using prior knowledge and statistical analysis.

\subsubsection{Waveform Features}

Handcrafted waveform features are obtained either by computing physical and statistical descriptors from the channel measurements \cite{guvencc2007nlos}, \cite{guvenc2007nloss}, \cite{marano2010nlos}, \cite{wymeersch2012machine}, or by directly using raw observation features output by devices or chips \cite{silva2016ir}, \cite{ferreira2021feature}, \cite{xie2025sora}.

Early studies set thresholds based on simple metrics such as noise level \cite{scholtz2002problems}, signal amplitude \cite{lee2002ranging}, and energy \cite{guvenc2005threshold}, and used them as decision criteria for ToA estimation under LoS/NLoS conditions. Guvenc et al. \cite{guvenc2005kurtosis} pointed out that feature selection based on a single threshold does not sufficiently reflect practical complex channels. They adopted kurtosis as a channel feature, arguing that it captures amplitude statistics and SNR information with low computational complexity, and significantly improves ToA estimation accuracy compared with \cite{guvenc2005threshold}. They introduced mean excess delay (MED) and root mean square (RMS) delay spread to characterize the delay properties of multipath channels \cite{guvencc2007nlos}, \cite{guvenc2007nloss}. Combining kurtosis with delay statistics characterizes LoS/NLoS channel differences from both amplitude and delay distributions. Later work \cite{marano2010nlos}, \cite{wymeersch2012machine} formed feature sets including received signal energy, maximum amplitude, rise time, MED, RMS delay spread, kurtosis, and range estimates, and analyzed the distribution differences of statistical features and the performance of different feature combinations under different channel propagation states. Zhou et al. \cite{zhou2015wifi} found through indoor measurements with commercial WiFi devices that signal skewness differs significantly across channel propagation states, and used skewness for channel state identification. 

Different from the above features that require additional computation, Silva et al. \cite{silva2016ir} evaluated simple DW1000-accessible features, including power difference, power ratio, and peak-to-lead delay, for LoS/NLoS identification in an industrial environment. However, compared with waveform-shape statistics such as kurtosis and skewness, these low-cost features showed weaker channel-state discriminability. Yu et al. \cite{yu2018novel} included SNR as a candidate feature and used the Pearson correlation coefficient (PCC) to analyze correlations between features, supporting more rational feature selection. Ferreira et al. \cite{ferreira2021feature} extracted raw features from the channel diagnostics of a commercial UWB transceiver, and evaluated different feature subsets with several machine learning models. They showed that a small number of low-overhead features can support real-time NLoS identification and mitigation. Different from studies using only statistical or raw features, Xie et al. \cite{xie2025sora} jointly considered the representation capability and acquisition cost of the two feature types, and used PCC to quantify the correlation between channel features and channel information, striking a favorable balance between feature expressiveness and system feasibility.

Waveform-oriented handcrafted features exhibit a trade-off between acquisition cost and representation capability \cite{xie2025sora}. Higher-order physical statistical features derived from the received waveform or CIR (e.g., maximum amplitude, MED, RMS delay spread, and kurtosis) exhibit strong physical interpretability. They can comprehensively characterize multipath structures, yielding better representation capability in environmental sensing. However, they incur relatively large computational overhead, require substantial hardware resources, and their distribution stability relies heavily on environmental conditions. Conversely, low-order lightweight raw features directly output by commercial devices during the ranging process (e.g., first-path amplitude, first-path index, and received signal strength) require no complex additional signal processing. They can be read directly from hardware registers or device interfaces, significantly reducing system latency and power consumption. Nevertheless, their information dimensionality and environmental representation capability are limited, making it difficult to fully capture the propagation characteristics of complex environments. Overall, these two feature types show a pronounced tradeoff between representation capability and acquisition cost, indicating that handcrafted features alone struggle to balance high accuracy, low overhead, and strong generalization in complex scenarios.

\subsubsection{Fingerprint Features}

Fingerprint features organize channel observations into location-discriminative representations that capture variations in observation patterns across spatial locations and support location matching, regression, or probabilistic inference.

At the outset, fingerprint features primarily relied on easily accessible RSS. The sample means of RSS from multiple BSs or access points (APs) were used to construct multidimensional fingerprint features for offline radio-map construction and online matching \cite{bahl2000radar}, \cite{kaemarungsi2004modeling}. Subsequently, Kaemarungsi et al. \cite{kaemarungsi2004properties} studied the statistical properties of RSS and analyzed the roles of the mean and variance in fingerprint features. Ibrahim et al. \cite{ibrahim2010cellsense} represented RSS fingerprints with probabilistic histograms, enabling probabilistic fingerprint matching. To address the sensitivity of RSS fingerprints to environmental dynamics and hardware heterogeneity, relative signal measurements, including signal strength ratios \cite{kjaergaard2008hyperbolic} and differences \cite{hossain2011ssd} between pairs of APs or BSs, were introduced to compensate for systematic RSS offsets. Lin et al. \cite{lin2016enhanced} exploited the common fluctuation trend of spatially adjacent locations and constructed neighborhood relative signal strength fingerprints from adjacent sampling points. Overall, RSS fingerprint features based on ratios and differences reduce their sensitivity to environmental dynamics and hardware heterogeneity.

Halperin et al. \cite{halperin2011tool} developed a CSI collection tool based on commercial WiFi devices, enabling fine-grained physical-layer channel information to be obtained and promoting research on CSI-based localization and sensing. Sen et al. \cite{sen2011precise} observed that CSI distributions at a given location retain statistically reproducible structures despite environmental dynamics. They exploited this reproducibility by fitting the measurements at each location with a Gaussian mixture model, using the cluster means and variances as fingerprints. To mitigate small-scale fading, Xiao et al. \cite{xiao2012fifs} used frequency and spatial diversity by dividing CSI into subchannels and spatially aggregating multi-antenna data, with the sum power over subchannels serving as the location feature. Wu et al. \cite{wu2013csi} applied time-domain filtering to suppress multipath reflections and computed effective CSI to fit a propagation model. For optimizing fingerprint dimensionality, Chapre et al. \cite{chapre2015csi} constructed composite fingerprints from amplitude and phase differences between adjacent subcarriers, improving fingerprint robustness by capturing the shape of CSI and adapting the fingerprint composition to environmental mobility. Rather than directly using or reorganizing CSI measurements, Chen et al. \cite{chen2019csi} modeled CSI with an autoregressive process and converted it into an information-entropy fingerprint. The resulting scalar feature has a compact form, stable statistical behavior, and strong location discriminability, and outperformed \cite{xiao2012fifs} and \cite{sen2011precise} in typical office environments. Building on \cite{chen2019csi}, \cite{chen2020aoa} extended fingerprint features from single CSI amplitude entropy to a dual-modal fingerprint combining amplitude entropy and phase AoA, enabling richer environmental information fusion.

Handcrafted fingerprint features enhance location-related information and suppress irrelevant disturbances through targeted statistical analysis and structural transformation of raw observations. However, their construction rules depend on heuristic experience rather than systematic learning. In complex and dynamic environments, such shallow handcrafted mappings have limited ability to capture intrinsic features that characterize channel propagation states and spatial relations.

\subsection{Transformation-Based Features}

Due to the intrinsic properties of observation modalities, multipath propagation, and environmental interference, propagation states and spatial geometric information cannot always be explicitly observed or extracted from raw observations. This motivates the use of domain transformation, signal analysis, and image encoding to represent channel observations. By transforming, reorganizing, or remapping raw measurements, transformation-based feature extraction methods explicitly reveal coupled, compressed, or masked location-related information, improving feature separability, stability, and learnability.

Raw observations describe wireless channels from only one or two dimensions among time, space, angle, and frequency, whereas location-related structures are embedded in multi-dimensional coupling. FT and IFT provide mathematical tools for multi-dimensional conversion of channel observations. Jin et al. \cite{jin2010indoor} used the IDFT to convert frequency-domain channel estimates into CIRs, and used the CIR amplitude as a high-dimensional fingerprint feature, which provides stronger location discriminability than RSS. Rather than constructing and matching fingerprint vectors, Wang et al. \cite{wang2019nlos} converted one-dimensional CIR into a two-dimensional impulse response spectrum using the short-time FT, where color encodes energy intensity. Chen et al. \cite{chen2017confi} organized continuously sampled CSI into a time-frequency matrix as a fingerprint feature image. By encoding signal features as images, these methods recast channel propagation state identification and position estimation as image classification problems, allowing classification, regression, or similarity matching with computer vision models. As CSI dimensionality increases, both the data structure and the carried information become more complex. In \cite{hejazi2021dyloc}, \cite{li2025d2}, raw CSI is mapped into structured angular-delay profile, turning high-dimensional complex observations that are difficult to interpret directly into visualized features reflecting multipath cluster distributions, user trajectories, and environmental disturbances. Such characterizations improve the physical interpretability of channel feature extraction and position estimation.

The wavelet transform (WT) has also been widely used for channel feature extraction and robust representation construction because it captures signal variations jointly in time and frequency. Unlike FT with its emphasis on global spectral analysis, WT describes global trends and local abrupt changes across multiple scales, making it suitable for nonstationary wireless channel measurements with pronounced local perturbations. Fang et al. \cite{fang2016channel} argued that \cite{xiao2012fifs}, \cite{kjaergaard2008hyperbolic}, and \cite{hossain2011ssd} used frequency and spatial diversity but lost channel variation and location information during fingerprint generation. To improve fingerprint stability, they performed multi-level discrete wavelet decomposition on CSI, preserving location information while suppressing severe fluctuations in raw CSI. Other studies noted that FT may remove useful information while filtering high-frequency noise \cite{wang2021indoor}, whereas WT supports CSI amplitude smoothing, denoising, feature extraction, and enhancement \cite{jiao2025robust}, \cite{rao2023mffaloc}. In addition, Cui et al. \cite{cui2020nlos} converted CIR into grayscale time-frequency images through the Morlet wavelet transform, while Wang et al. \cite{wang2022multi} applied one-dimensional wavelet packet decomposition to thresholded CIR, constructed an energy decomposition tree, then converted it into color coefficient images. Experimental results show that multi-scale time-frequency representations preserve local discriminative patterns, strengthening feature learning and propagation state identification.

\begin{figure}[t]
\centerline{\includegraphics[width=0.8\linewidth]{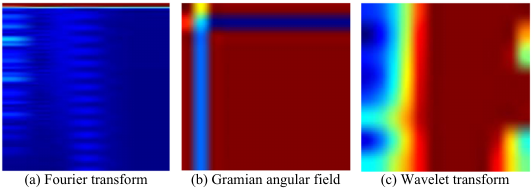}}
\caption{1D signals conversion to 2D images \cite{kerdjidj2024exploiting}.}
\vspace{-2mm}
\label{fig:transformation_features}
\end{figure}

Image encoding and graph reconstruction have also been used to uncover implicit patterns in wireless measurements. A common strategy maps one-dimensional signals to two-dimensional images such as those in Fig.~\ref{fig:transformation_features}. A common strategy maps one-dimensional signals to two-dimensional images. For example, the Gramian angular field (GAF) uses polar-coordinate mapping and Gramian matrix construction to convert one-dimensional CIR into two-dimensional image textures, preserving temporal correlations in the image domain while increasing the distinguishability of the overall structure \cite{wang2015imaging}, \cite{deng2023uwb}, \cite{kerdjidj2024exploiting}. Another strategy builds graphs according to similarity or adjacency relations among measurements, reference points, and APs, then uses graph features or embeddings to enhance modeling of local associations and global structures. Liu et al. \cite{liu2022device} mapped CSI subcarrier sequences into a limited penetrable horizontal visibility graph, preserving inter-carrier correlations through graph topology. Zhang et al. \cite{zhang2023domain} constructed a heterogeneous directed graph with reference points and APs as nodes, using RSSI relations as edge features and graph convolution to obtain more discriminative structured embeddings. This design alleviates label scarcity and domain shift in crowdsourcing scenarios. \cite{liu2025graph} used APs as graph nodes, fine-grained CSI fingerprints as node features, and inverse AP distances as edge weights to construct an undirected graph. This design explicitly characterizes the spatial topology of fingerprint data, with temporal modeling further extracting spatiotemporal correlation features.

\subsection{Implicit Representation Learning}

Implicit representation learning exploits the nonlinear mapping capability of neural networks and uses position labels, propagation states, and spatio-temporal correlations as constraint signals to learn representations related to signal propagation and spatial location directly from channel observations. These representations do not necessarily correspond to observable channel parameters or explicit features with well-defined physical meanings. Instead, they are encoded as intermediate-layer embeddings, hidden states, or low-dimensional representations that implicitly characterize channel propagation states and spatial geometric properties.

The most direct form is end-to-end mapping from raw observations to positions under the supervision of ground-truth positions. Hsieh et al. \cite{hsieh2019deep} divided indoor areas into spatial blocks and trained multilayer perceptron (MLP) and convolutional neural network (CNN) models using raw RSS and CSI, respectively. Subsequently, a range of CNN architectures and variants were applied to end-to-end mapping from fingerprints to positions \cite{chin2020intelligent}, \cite{cerar2021improving}. Wang et al. \cite{wang2025learning} selected frequently observed APs to construct fixed-length continuous RSSI sequences, with ResNet-18 used as the feature extraction network. In this way, supervision constrains not only single-point samples but also continuity in observation sequences. Zhang et al. \cite{zhang2023csi} argued that CNNs are inefficient in extracting global contextual information from high-dimensional observations because of limited convolutional depth and receptive field. They combined attention mechanisms with residual structures to capture local details and global context in CSI, improving the discriminability of high-dimensional channel representations and their adaptability to complex environments.

Supervision signals also extend to channel state labels and geometric measurements that are directly related to propagation mechanisms. In \cite{bregar2018improving}, \cite{jiang2020uwb}, and \cite{angarano2021robust}, raw CIR sequences were fed directly into CNNs, with channel states or ranging values used as supervision for representation learning. Jiang et al. \cite{jiang2020uwbnlos} and Yang et al. \cite{yang2022robust} combined CNNs with long short-term memory (LSTM) networks to extract spatio-temporal features from CIR or RSS sequences. CNNs are used to capture local signal patterns, while the LSTM models long-range dependencies through cell states and gating mechanisms.

Propagation mechanisms have also been introduced as explicit constraints, allowing implicit representations to remain effective for task labels while being consistent with physical propagation laws. Olivares et al. \cite{olivares2021applications} developed a physics-informed neural network (PINN) for WiFi signal propagation prediction by encoding geometric relations, material properties, and integral path effects in electromagnetic propagation as channel information, improving the efficiency of signal distribution modeling in complex indoor scenarios. Lombardi et al. \cite{lombardi2025reducing} incorporated signal propagation partial differential equations into the PINN loss as physical constraints.

\subsection{Channel Charting-Based Representation Learning}

\begin{figure}[t]
\centerline{\includegraphics[width=0.7\linewidth]{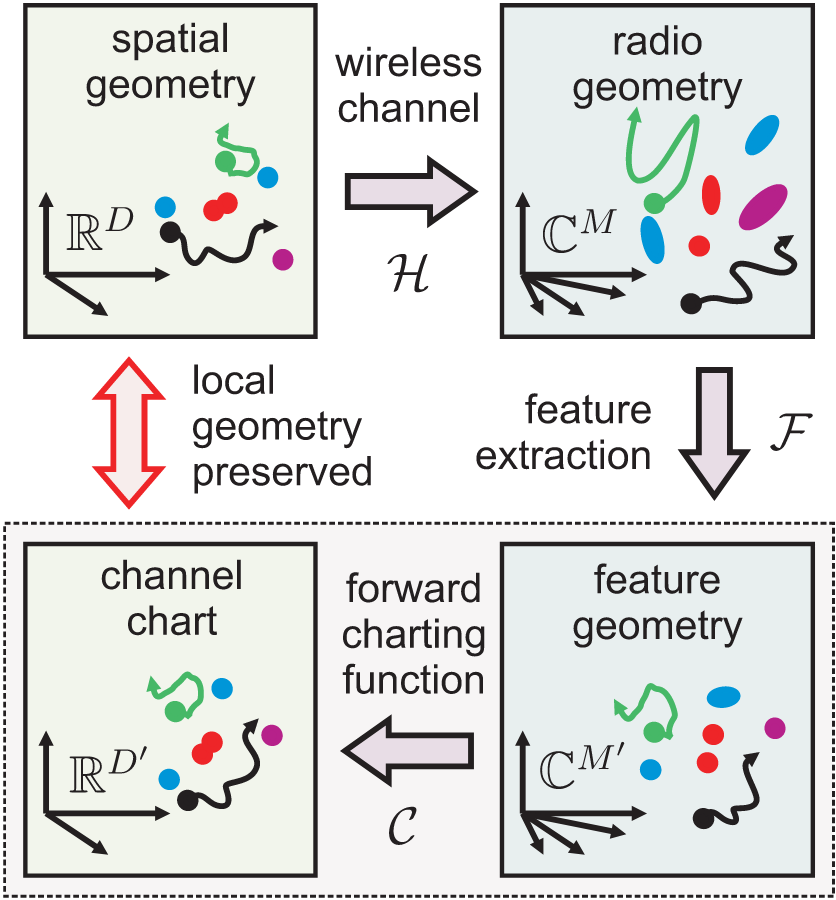}}
\caption{Summary of the geometries involved in CC \cite{studer2018channel}.}
\vspace{-2mm}
\label{fig:cc_studer2018channel}
\end{figure}

Channel charting (CC)-based representation learning derives low-dimensional embeddings from high-dimensional channel observations to reflect relative spatial relations. Studer et al. \cite{studer2018channel} pioneered the CC framework based on the assumption that physical spatial proximity leads to similar channel propagation characteristics, and that similarities in the observation space should be preserved in the low-dimensional representation space. With appropriate similarity metrics and dimensionality reduction methods, local neighborhood relations, trajectory continuity, and geometric structures hidden in high-dimensional observations become explicit in the low-dimensional space, as shown in Fig.~\ref{fig:cc_studer2018channel}.

Subsequently, Huang et al. \cite{huang2019improving} introduced representation constraints within the autoencoder latent space, encouraging low-dimensional embeddings to maintain geometric consistency while reconstructing channel observations. Lei et al. \cite{lei2019siamese} leveraged Siamese neural networks (SNNs) to parameterize the nonparametric Sammon mapping as a learnable model, thereby enabling the network to learn channel charts of low dimensionality from distance relationships between channel sample pairs. Ferrand et al. \cite{ferrand2021triplet} constructed weakly supervised signals using triplet samples, driving the low-dimensional embeddings to preserve proximity relations and trajectory continuity in the sampling process.

Wireless propagation characteristics and relative distances have also been incorporated into similarity or dissimilarity metrics to make low-dimensional representations more faithful to the spatial structure of channel variations. Le et al. \cite{le2021efficient} designed a cosine distance metric insensitive to small-scale fading and phase rotation, and combined the resulting distance matrix with Isomap to improve the local geometric quality of channel charts. Stahlke et al. \cite{stahlke2023indoor} used delay information in CIR to construct a time distance metric with a linear relation to global physical distance, strengthening the preservation of global geometric structure in channel charts. Stephan et al. \cite{stephan2024angle} transformed CIR into the angle-delay domain and designed a dissimilarity metric from angle and delay information, with side information such as timestamps incorporated to improve local and global structures. Ahadi et al. \cite{ahadi2025tdoa} introduced geometric constraints, including time difference of arrival (TDoA), transmitter and receiver positions, and relative displacement. These constraints together with NLoS mitigation enhance the robustness of channel chart representations under complex propagation conditions.

\subsection{Channel Knowledge Map-Based Representation Learning}

\begin{figure}[t]
\centerline{\includegraphics[width=0.7\linewidth]{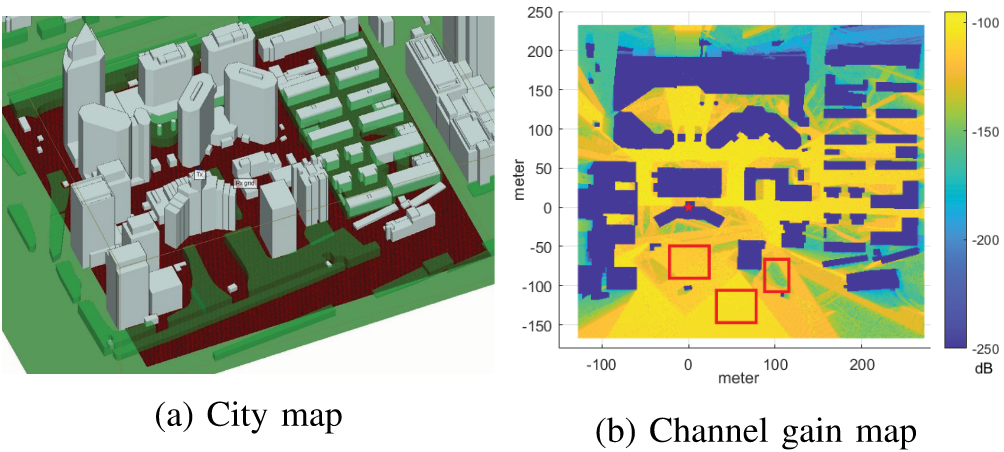}}
\caption{City map and corresponding channel gain map \cite{xu2024much}.}
\vspace{-2mm}
\label{fig:ckm_xu2024much}
\end{figure}

The channel knowledge map (CKM) is an environment-aware framework for channel representation and can also be understood as a prior digital map constructed for a specific area. It characterizes the spatial distribution of wireless channels while treating the wireless propagation environment as an explicitly modelable entity. Rather than simply storing static environmental elements such as buildings or vegetation, CKM indexes channel knowledge by positions or spatial regions and stores or predicts the corresponding channel knowledge, with the channel gain map in Fig.~\ref{fig:ckm_xu2024much} as an example. Specifically, CKM incorporates environmental maps, ray-tracing results, and historical measurements to pre-compute, learn, or recover channel features at specific locations, forming a database with rich channel knowledge \cite{zeng2024tutorial}, \cite{ren2026channel}, \cite{zeng2021toward}. Unlike fingerprint databases that store mappings between signals and positions, or CC methods that learn unsupervised embeddings between signals and relative positions, CKM emphasizes causal physical relations between the environment and the channel. It establishes physically meaningful associations among position, environment, and channel features, making channel knowledge queryable, recoverable, and reusable.

Zeng et al. \cite{zeng2021toward} introduced the CKM concept and defined it as a database for a specific area, where location information such as transmitter and receiver positions serves as the index for the corresponding channel knowledge. With the development of CKM, channel knowledge has been instantiated through different physical observables or system parameters, forming concrete CKM variants that support channel feature extraction at different granularities. Wu et al. \cite{wu2021environment} proposed the channel path map (CPM) and beam index map (BIM), using multipath parameters and optimal beam indices as channel knowledge, respectively. CPM stores location-related multipath parameters and predicts channel path information for all potential user locations in the coverage area. BIM directly stores the optimal beam-pair index corresponding to each location and does not require explicit channel parameter estimation. Li et al. \cite{li2022channel} constructed a channel gain map (CGM) using the expectation maximization (EM) algorithm. Channel gains are modeled as a Gaussian mixture distribution, where different Gaussian components describe statistical variations and EM estimates the model parameters. Wu et al. \cite{wu2024environment} integrated dynamic sensing information into CKM-assisted channel estimation, decomposing the environment into quasi-static and dynamic components and using a channel angle map (CAM) to characterize their angle information.

In practical systems, channel measurements are collected only at finite and unevenly distributed sampling points, making CKM naturally discrete and sparse. Therefore, recovering channel features in unmeasured areas relies on spatial channel correlation, where measured features such as path loss, channel gain, and delay spread are extended to nearby regions through interpolation, matrix completion, or tensor decomposition. Sun et al. \cite{sun2022propagation} combined interpolation with matrix completion to supplement sparse observations and exploit the low-rank structure of propagation maps, capturing both local spatial correlation and global structure. For spectrum map construction, they combined interpolation with tensor decomposition to extract more complete spectrum features across spatial and frequency dimensions \cite{sun2024integrated}. Xu et al. \cite{xu2024much} discussed data requirements in CGM construction and showed that CKM recovery accuracy is affected by sampling density, spatial correlation, and propagation environment complexity. 

Spatial smoothness alone is insufficient to characterize propagation variations in complex environments, motivating the incorporation of environmental structures and propagation mechanisms into CKM feature extraction. Chen et al. \cite{chen2024diffraction} used geometry-model-assisted deep learning to jointly recover radio maps and virtual environmental structures from RSS observations, with diffraction and scattering effects explicitly considered. CKMImageNet \cite{wu2025ckmimagenet} combines position-labeled channel knowledge, high-fidelity environmental maps, and visual representations, providing a data foundation for models to learn the relation between spatial structures and channel features from environmental images. Fu et al. \cite{fu2025ckmdiff} treat sparse or degraded CKMs as incomplete channel feature maps and use conditional diffusion models to restore missing or noise-affected channel knowledge. Related variants incorporate physical constraints \cite{zhu2025channel}, beam conditions \cite{zhao2026beamckmdiff}, and flow matching mechanisms \cite{huang2026channel} into the generation process to improve the physical consistency and construction efficiency of recovered channel features. These methods extract more complete channel features from environmental layout, propagation blockage, and generative priors, extending CKM toward more complex spatial propagation structures.

When CKM evolves from discrete grids to continuous spatial representations, channel features are no longer only values at sampling points or grid cells, but are treated as fields that vary continuously with position, direction, and frequency. This view provides finer-grained propagation priors and supports the recovery and query of channel features in continuous space. RF-3DGS \cite{zhang2026rf} models wireless propagation as a renderable RF radiance field and uses 3D Gaussian splatting to recover spatial channel states from sparse samples. Zhou et al. \cite{zhou2026f} proposed F$^4$-CKM, which learns a CKM with spatial and frequency awareness through wireless radiance field rendering and predicts channel features in continuous spatial and frequency domains. BiWGS \cite{zhou20256d} considers dynamic transmitter and receiver positions, extending conventional 2D/3D CKM to 6D to characterize channel variations under different transmitter-receiver location combinations.

\subsection{Self-Supervised Representation Learning}

Self-supervised representation learning treats the internal structure and consistency within channel observations as self-supervisory signals. By designing pretext tasks such as signal reconstruction, masked recovery, and contrastive learning, it learns channel features from large amounts of unlabeled data that preserve latent propagation structures and support cross-scenario transfer and downstream adaptation \cite{yang2025revolutionizing}. According to the design of pretext tasks, self-supervised learning (SSL) mainly follows generative and contrastive modes.

\subsubsection{Generative SSL Representation Learning}

Generative SSL learns structural information and contextual relations associated with channel propagation from raw wireless measurements through signal reconstruction, masked recovery, and data generation. These methods either learn low-dimensional representations through reconstruction or masked recovery, or model the underlying data distribution to synthesize virtual samples for representation learning.

The encoder-decoder architecture provides a natural way to learn compact and reconstructable channel representations from measurements. In this framework, the encoder compresses high-dimensional observations into a low-dimensional latent space, while the decoder reconstructs the original signal from the latent representation. The latent space is limited in dimensionality, and reconstruction or recovery tasks require key input information to be preserved as much as possible. These constraints guide the model to retain the structural factors that contribute most to signal reconstruction, leading to compression \cite{hinton2006reducing} and denoising \cite{vincent2010stacked} of raw observations.

Pioneering studies interpreted communication systems as end-to-end compression and reconstruction tasks under an autoencoder (AE) structure. The learned features outperformed finely modeled schemes and showed stronger robustness to nonlinearities and system imperfections \cite{o2017introduction}, \cite{felix2018ofdm}. To reduce the CSI feedback overhead caused by multiple antennas in MIMO systems, Wen et al. \cite{wen2018deep} proposed a CNN-based autoencoder to learn spatio-temporal channel relations, achieving high-quality reconstruction under low compression ratios. Related work further explored temporal correlation \cite{wang2018deepLSTM}, multi-resolution design \cite{lu2020multi}, distributed structures \cite{mashhadi2020distributed}, and attention mechanisms \cite{mourya2022spatially}, \cite{cui2022transnet}. It has been applied to channel estimation and recovery \cite{kang2018deep}, \cite{yi2020deep}, channel prediction \cite{yang2025wirelessgpt}, \cite{liu2025wifo}, and beamforming \cite{alkhateeb2018deep}. These works show that low-dimensional latent variables learned by encoder-decoder models preserve dominant propagation structures and correlations rather than merely compressing the input.

The same architecture has been extended to wireless sensing and localization. \cite{khatab2017fingerprint} and \cite{kim2018scalable} used AE-based RSS reconstruction constraints to extract more compact and robust fingerprint features. Unlike the deterministic reconstruction of AE, the variational autoencoder (VAE) maps data into a continuous probability distribution and guides the model toward smooth, continuous, and generalizable latent representations. Stahlke et al. \cite{stahlke2021estimating} used a VAE to model CIR distributions under LoS conditions with reliable ToF estimates, and constructed anomaly scores from the similarity between observed and reconstructed CIRs for NLoS identification and ranging enhancement. Kim et al. \cite{kim2021multiview} treated CSI from different AP groups as multiview inputs, and used latent-variable modeling with dominant-view classification to fuse complementary views while suppressing uninformative ones, improving the robustness of fingerprint representations. Li et al. \cite{li2023variational} used variational learning to jointly extract distance-related and environment-related information from wireless signals, supporting concurrent distance estimation and environmental identification. In addition, several studies use VAE to learn latent distributions of real channel features or wireless fingerprints, then synthesize virtual samples consistent with the original distributions to enrich training data, reduce manual sampling cost, and improve robustness under sample scarcity and domain shifts \cite{chen2020fido}, \cite{chen2022fidora}.

Most encoder-decoder methods rely on convolutional or fully connected structures, with modeling capacity mainly concentrated on local features. In large-scale arrays and wideband systems, high-dimensional channel observations require long-range dependency modeling across subcarriers, antennas, and time dimensions, which remains difficult for such structures even with larger receptive fields or deeper networks. For this reason, Transformer architectures have been introduced into generative representation learning. Cui et al. \cite{cui2022transnet} proposed TransNet based on the Transformer architecture. By introducing attention mechanisms, it learns correlations among different parts of CSI and preserves more global structural information under limited compression ratios.

The masked autoencoder (MAE) represents an important masked-recovery pretext task for generative SSL in wireless research. He et al. \cite{he2022masked} showed that reconstructing masked content from partially visible observations can drive the model to learn stable latent representations. Following this idea, Liu et al. \cite{liu2025wifo} proposed a spatio-temporal-frequency wireless foundation model for channel prediction. The model partitions three-dimensional CSI into 3D patches and designs random masking, temporal masking, and frequency masking tasks, enabling one model to handle prediction in both the time and frequency domains. Alikhani et al. \cite{alikhani2025lwm} pretrained a Transformer encoder on large-scale wireless channel data with MAE and learned task-agnostic transferable channel embeddings for downstream wireless sensing and communication tasks.

Another generative path relies on the generative adversarial network (GAN), which fits wireless signal or fingerprint sample distributions through adversarial training between a generator and a discriminator. The generator learns the latent sample distribution, whereas intermediate discriminator layers implicitly encode discriminative features between real and generated samples. In existing studies, the value of GANs lies less in directly producing interpretable channel representations than in expanding fingerprint databases \cite{li2019af}, augmenting data \cite{nkrow2023transfer}, \cite{zhang2023learning}, and completing radio maps \cite{yoon2024gan}.

\subsubsection{Contrastive SSL Representation Learning}

Contrastive SSL learns discriminative channel representations from raw wireless measurements by constructing sample consistency or discrepancy constraints. Focusing on the relative relations among observations, this approach optimizes the latent space structure by drawing the embeddings of semantically similar samples closer while pushing unrelated ones apart. Ultimately, this structural optimization suppresses irrelevant factors (e.g., device heterogeneity, system bandwidth, and instantaneous perturbations), yielding highly robust semantic representations for cross-scenario transfer and downstream tasks.

Xu et al. \cite{xu2023channel} assumed that neighboring positions have similar CSI similarity metrics and used this relation to construct positive and negative sample pairs, mapping CSI from adjacent positions to similar feature representations for channel estimation. Salihu et al. \cite{salihu2024self} built paired samples from augmented views of the same CSI and neighboring subcarriers, using their consistency to learn macro- and micro-fading representations that are less sensitive to fading variations and system impairments. Jiang et al. \cite{jiang2025mimo} proposed a cross-modal contrastive learning model that aligns CSI-CIR pairs in embedding space, achieving joint learning of time-domain and frequency-domain features without data augmentation.

Unlike approaches relying primarily on observation consistency for contrastive constraints, Ozeki et al. \cite{ozeki2025cellular} incorporated task labels into contrastive learning. For indoor localization with cellular signals, they borrowed sequence modeling ideas from large language models (LLMs). By treating RSS values as words and their spatial distributions as sentences, they encoded RSS measurements together with the corresponding BS IDs into token sequences for an adapted GPT-2. By computing similarities between positive and negative samples, this method drives the learned representations to accurately reflect spatial proximity relations.

\subsubsection{Hybrid SSL Representation Learning}
Recent studies integrate reconstruction constraints from generative SSL with relational constraints from contrastive SSL to simultaneously learn structured and discriminative channel representations. Guler et al. \cite{guler2025multi} combined masked reconstruction with masked contrastive learning, using noise, fading, and partial observability as natural augmentation sources to contrast different masked views of the same CSI and learn representations with stronger structural coherence and separability. Inspired by information bottleneck theory, Pan et al. \cite{pan2025large} jointly optimized spatial-frequency masked channel modeling, domain-transform invariance, and location-invariant contrastive learning, forming a hybrid representation that preserves channel structure while suppressing task-irrelevant variations. Jiao et al. \cite{jiao2025addressing} treated electromagnetic propagation as the physical consistency basis between environmental and channel modalities, aligning environment-channel pairs and using CSI reconstruction to constrain path-information completeness for multi-modal channel representation learning. 

\begin{table*}[t]
\caption{Comparison of channel feature extraction and representation learning methods.}
\label{tab:feature_representation_learning}
\centering
\scriptsize
\renewcommand{\arraystretch}{1.2}
\setlength{\tabcolsep}{3.5pt}
\begin{lrbox}{\featuretablebox}
\begin{tabular}{p{0.23\textwidth}|c|c|c|c|p{0.27\textwidth}}
\hline
\multicolumn{1}{c|}{\multirow{2}{*}{\textbf{Method}}} & \multicolumn{2}{c|}{\textbf{Representation Capability}} & \multicolumn{1}{c|}{\multirow{2}{*}{\textbf{Generalization Capability}}} & \multicolumn{1}{c|}{\multirow{2}{*}{\textbf{Labeled Data}}} & \multicolumn{1}{c}{\multirow{2}{*}{\textbf{References}}} \\
\cline{2-3}
\multicolumn{1}{c|}{} & \textbf{Granularity} & \textbf{Expressiveness} & \multicolumn{1}{c|}{} & \multicolumn{1}{c|}{} & \multicolumn{1}{c}{} \\
\hline\hline
Handcrafted Features & \lowstar & \lowtri & \lowcirc & \lowsq & \cite{guvencc2007nlos} \cite{marano2010nlos} \cite{wymeersch2012machine} \cite{ferreira2021feature} \cite{xie2025sora} \cite{xiao2012fifs} \cite{chapre2015csi} \cite{chen2019csi} \cite{chen2020aoa} \\
Transformation-Based Features & \medstar & \medtri & \medcirc & $\times$ & \cite{jin2010indoor} \cite{chen2017confi} \cite{hejazi2021dyloc} \cite{fang2016channel} \cite{cui2020nlos} \cite{deng2023uwb} \cite{liu2022device} \\
Implicit Representation Learning  & \medstar & \hightri & \lowcirc & \highsq & \cite{hsieh2019deep} \cite{zhang2023csi} \cite{bregar2018improving} \cite{yang2022robust} \cite{fontaine2023transfer} \cite{olivares2021applications} \\
CC-Based Representation Learning  & \medstar & \lowtri & \lowcirc & $\times$ & \cite{studer2018channel} \cite{huang2019improving} \cite{lei2019siamese} \cite{ferrand2021triplet} \cite{le2021efficient} \cite{stephan2024angle} \\
CKM-Based Representation Learning  & \highstar & \medtri & \lowcirc & \highsq & \cite{zeng2021toward} \cite{wu2021environment} \cite{li2022channel} \cite{wu2024environment} \cite{chen2024diffraction} \cite{sun2022propagation} \cite{zhang2026rf} \cite{zhou20256d}\\
Self-Supervised Representation Learning  & \highstar & \hightri & \highcirc & $\times$ & \cite{wen2018deep} \cite{stahlke2021estimating} \cite{liu2025wifo} \cite{salihu2024self} \cite{guler2025multi} \cite{pan2025large}\\  \hline
\end{tabular}
\end{lrbox}
\usebox{\featuretablebox}
\par\vspace{2pt}
\parbox{\wd\featuretablebox}{\raggedright\scriptsize\textbf{\textit{Note:}} One, three, and five symbols indicate low, medium, and high levels, respectively; \checkmark\ and $\times$ indicate yes and no.}
\end{table*}

{
\subsection{Roles of Channel Representations}

These studies show that channel representation is neither simply a compression or reconstruction of the observation, nor does it require fully reconstructing the channel response or explicitly identifying every propagation path. Instead, it emphasizes the learning-driven organization and refinement of spatial propagation structures in channel measurements, transforming complex channel observations into structured and discriminative location-aware representations. Its main roles can be summarized as follows.

\begin{enumerate}
\item Based on channel statistics or physical propagation laws, channel representation extracts delay, angle, path gain, and multipath structure information from raw measurements, capturing physical propagation properties, and correlations across time, space, and frequency.
\item High-dimensional channel information is mapped into structured and discriminative feature embeddings, improving separability and continuity in the feature space.
\item The effects of environmental dynamics, hardware differences, and noise are suppressed or separated, which strengthens feature stability and generalization.
\item It maps measurements collected under heterogeneous conditions, such as different wireless technologies, system configurations, and device parameters, into shared or comparable feature spaces, thereby facilitating multi-source information fusion.
\end{enumerate}

These methods organize channel representations in distinct ways and preserve different aspects of channel information. We have summarized the characteristics of the channel feature extraction and representation learning methods in Table~\ref{tab:feature_representation_learning} for comparison. However, in practice, the choice of channel feature extraction and representation learning methods depends on the specific task, performance requirements, data conditions, and implementation cost.

Fundamentally, channel representation aims to extract location-related information from raw observations while suppressing irrelevant variations caused by environmental dynamics, hardware differences, and system configurations. Consequently, channel representation is no longer merely a conventional black-box input transformation or feature preprocessing step. Rather, it serves as a critical link between channel observations and position estimation models, providing an essential foundation for understanding and overcoming the challenges of high-accuracy localization.}

\section{Channel Representation-Based Localization}
\label{sec:representation_based_localization}
Wireless channel observations inherently correlate with spatial locations. However, this relationship is highly intricate and can couple location information with extraneous factors from devices, systems, or environments. The quality of channel representations and their subsequent application directly determine the overall performance of localization models. We systematically review existing methods from the perspectives of representation acquisition and organization, comparing their differences in localization accuracy, data requirements, applicable conditions, and generalization capabilities. 

\subsection{Classical Representation-Based Localization}

\begin{figure}[t]
\centerline{\includegraphics[width=0.9\linewidth]{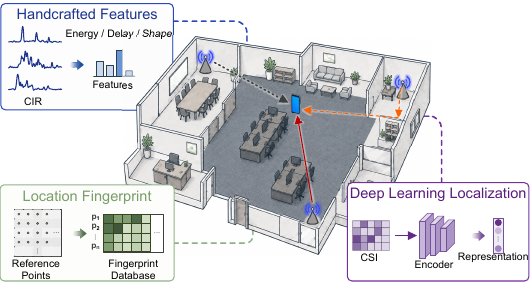}}
\caption{Representative realizations of localization with classical channel representations, covering handcrafted waveform features, channel fingerprints, and supervised deep representation learning.}
\vspace{-2mm}
\label{fig:classical_representation}
\end{figure}

Classical representation-based localization adopts task-oriented feature extraction and representation learning, without a dedicated design for representation structure or cross-scenario consistency. These methods assume that data from the training and test scenarios are independent and identically distributed. Representations extracted or learned from raw channel observations are expected to preserve sufficient location-related information for direct use in location inference. Typical realizations in Fig.~\ref{fig:classical_representation} include waveform feature extraction, location fingerprints, and implicit representations learned through fully supervised deep networks.

Multipath and NLoS propagation change the energy distribution and delay structure of the CIR, inducing positive biases in ToF/ranging estimations that rely on the direct path, and causing time-domain response waveforms to exhibit distinctly different distribution characteristics across various propagation states. Earlier work identified NLoS conditions by extracting statistical features from the CIR, mitigating localization errors induced by NLoS by assigning lower weights, discarding, or correcting measurement errors. Guvenc et al. \cite{guvencc2007nlos} extracted multipath statistics such as kurtosis, MED, and RMS delay spread from UWB channels, and used a joint likelihood ratio test to identify LoS/NLoS states and assign smaller WLS weights to measurements likely affected by NLoS bias. Related work extracted channel features such as received signal energy, maximum amplitude, rise time, MED, RMS delay spread, and kurtosis from the CIR for ranging error compensation through support vector machines (SVMs) \cite{marano2010nlos}, \cite{wymeersch2012machine} and Gaussian regression models \cite{wymeersch2012machine}. Ferreira et al. \cite{ferreira2021feature} used low-overhead features directly available from commercial UWB chips to select feature combinations for NLoS identification and ranging error mitigation with machine learning models. These methods have explicit physical meanings and easily integrate with geometric computation, improving ranging and localization performance. However, handcrafted features struggle to adequately describe fine-grained multipath discrepancies in complex environments, exhibiting limited expressive capacity for channel propagation. Concurrently, feature distributions shift due to environmental variations, thereby restricting generalization capabilities.

Encoding position-related channel discrepancies as location fingerprints and performing location inference through online matching or regression is another realization strategy. The RADAR system proposed by Bahl and Padmanabhan \cite{bahl2000radar} assumes that received signal distributions at distinct reference points are distinguishable. It constructs an RSS fingerprint map offline and performs nearest-neighbor matching online for indoor position estimation. With CSI becoming available from commercial WiFi network interface cards, researchers began to replace RSS with CSI for finer-grained fingerprint construction. Xiao et al. \cite{xiao2012fifs} proposed FIFS, which exploits the spatial separability of multi-subcarrier CSI amplitudes to construct location fingerprints and uses probabilistic matching to achieve higher localization resolution than RSS. Chen et al. \cite{chen2020aoa} used autoregressively modeled CSI amplitude entropy and phase-derived AoA as a composite fingerprint. A bivariate kernel regression scheme was then used for location inference, achieving decimeter-level localization accuracy in complex indoor environments. These methods map discriminative channel fingerprints to spatial positions and can achieve high accuracy under consistent training and test conditions. However, their reliance on scenario, device, and system consistency leads to high data collection and fingerprint maintenance costs, while environmental or deployment changes often cause significant performance degradation.

The introduction of deep networks enables researchers to directly learn representations relevant to locations from measurements, achieving end-to-end geometric information or position estimation. Bregar et al. \cite{bregar2018improving} fed raw CIR sequences into a CNN to extract local waveform structures for ranging error compensation. Different from CNNs that focus on local pattern extraction, Kim et al. \cite{kim2022uwb} used LSTM to model CIR sequences and classified NLoS-induced error levels, which were then used to correct distance observations and noise covariances in an EKF-based localization process. Yang et al. \cite{yang2022robust} used RSS and distance measurements as sequential inputs, with CNNs extracting local spatial features from anchor observations and LSTMs modeling temporal correlations for 3D position estimation. Zhang et al. \cite{zhang2023csi} introduced attention into residual CNN-based CSI fingerprint learning, strengthening global contextual modeling beyond local receptive fields and producing more discriminative deep fingerprints. These methods avoid explicit handcrafted feature design and learn location-related patterns automatically from high-dimensional channel observations. By continuously enhancing representation capabilities through model design and optimization, these methods achieve higher localization accuracy. However, they require large numbers of labeled samples and rarely consider environmental adaptation or model generalization, so cross-scenario deployment faces evident performance degradation.

\begin{figure}[t]
\centerline{\includegraphics[width=0.9\linewidth]{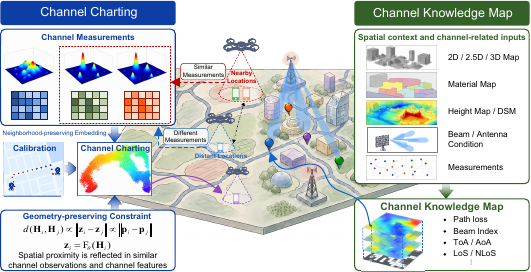}}
\caption{Spatial organization of channel representations through channel charting and CKM, where relative sample geometry and position-indexed channel knowledge support location inference.}
\label{fig:spatial_structure}
\end{figure}

\subsection{Spatial Structure-Based Localization}

Although the localization result from a single channel measurement is ultimately a coordinate, channel measurements at different locations are not isolated from each other. Their variations across physical locations and mutual relations reflect the underlying spatial geometry. Spatial structure-based localization exploits this property and treats localization as the inverse problem of the channel measurement process. The position is determined within a spatial structure recovered from channel measurements or constructed in advance. Existing studies have developed two main spatially structured localization frameworks, comprising CC and CKM. CC focuses on utilizing relationships between channel samples to recover relative spatial structures, whereas CKM leverages channel knowledge indexed by positions to describe the channel distribution across space. As shown in Fig.~\ref{fig:spatial_structure}, both frameworks embed channel representations into spatial structures, providing essential support for subsequent location inference.

\subsubsection{Channel Charting}
The channel chart obtained from channel measurements only provides relative relationships within space, reflecting spatial associations between samples without directly yielding position estimation in an actual coordinate system. Therefore, research on CC-based localization has gradually shifted from the recovery of relative spatial structure to the calibration of relative channel charts into usable location representations. The calibration step in Fig.~\ref{fig:spatial_structure} aligns channel charts with physical coordinates, and existing studies have used anchors, side information, or geometric constraints to establish this alignment. Huang et al. \cite{huang2019improving} constrained channel charts with side information, ensuring that low-dimensional embeddings preserve channel similarity while aligning closely with the true spatial structure, and acquiring approximate localization capability. 

\begin{figure}[t]
\centerline{\includegraphics[width=\linewidth]{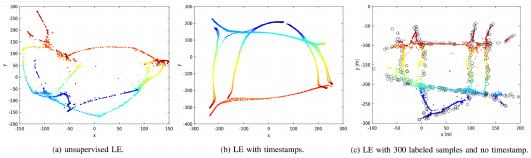}}
\caption{Visualization of the channel charts from \cite{deng2021network}.}
\label{fig:cc_calibrated}
\end{figure}

Observations of the same user from multiple APs or BSs provide richer spatial constraints, enabling channel charts to no longer rely entirely on relative relationships from a single view, which alleviates issues concerning scale, orientation, and local ambiguities. Pihlajasalo et al. \cite{pihlajasalo2020absolute} constructed unsupervised multipoint channel charts using multiple BSs, and mapped the combined channel charts to an absolute coordinate system through affine or conformal transformations. This work provides a route for CC to move from relative representation toward absolute localization. Deng et al. \cite{deng2021network} proposed a semi-supervised multipoint CC-based localization method that jointly incorporates the CSI, timestamps, channel quality, and a limited number of location labels from distributed base stations for graph construction and manifold constraints. Fig.~\ref{fig:cc_calibrated} shows the channel chart aligned with physical coordinates using only 10\% labeled CSI samples, and the method achieved meter-level localization accuracy. This result indicates the value of CC as a low-label spatial-structure modeling method that significantly reduces the training cost of localization systems when labeled samples are limited.

To make channel charts calibrated by a few anchors can be stably utilized for localization, subsequent studies paid more attention to whether the distance scale in the channel chart is consistent with physical space. Stahlke et al. \cite{stahlke2023indoor} used a simple linear transformation to map relative channel charts to the real coordinate system and realized CC-assisted fingerprint localization with only a limited number of reference coordinate points (10--20 points). In related work, Stephan et al. \cite{stephan2024angle} evaluated the ability of CC to support real position estimation through the absolute localization error after affine coordinate transformation. These results show that channel charts possessing a robust global structure after minimal coordinate calibration can significantly reduce the dependence on position labels and achieve submeter-level localization accuracy in indoor scenarios.

Some works have extended their focus to enabling CC to directly serve position estimation in real coordinate systems. Taner et al. \cite{taner2023channel}, \cite{taner2025channelcharting} attempted to constrain channel charts to real-world coordinate systems, turning channel charts from relative embeddings under arbitrary coordinates into representations that are more directly usable for coordinate estimation. Stephan et al. \cite{stephan2025three} extended CC to 3D indoor localization and multi-floor scenarios, addressing the vertical dimension and floor ambiguity that commonly appear in practical environments. Poeggel et al. \cite{poeggel2025passive} applied CC to passive target localization, showing that the idea of CC also extends from active terminal localization to device-free spatial sensing.

CC provides a low-label and structured path for channel representation in localization. It no longer relies entirely on densely labeled fingerprints. Instead, CC recovers spatial structure from relative relations between channel observations, then achieves location inference through a few anchors, side information, or coordinate transformations. Such methods have distinct advantages in scenarios where labeled samples are scarce and infrastructure can provide continuous channel measurements. However, channel charts also require observation data that adequately cover spatial structure. Meanwhile, they exhibit different degrees of distortion compared to actual space, presenting inherent challenges in global spatial-structure recovery and absolute coordinate transformation. When the environment changes, channel features shift accordingly, which weakens model transferability. Collectively, these problems limit the practical application of this class of algorithms.

\subsubsection{Channel Knowledge Map}
Spatial structure is also integrated into the localization process through channel knowledge. CKM condenses propagation characteristics into position-indexed channel knowledge as in Fig.~\ref{fig:spatial_structure}, enabling new channel observations to be interpreted, matched, and linked back to corresponding locations in an existing CKM. In this sense, CKM-based localization is viewed as an inverse problem of channel observations. Zeng et al. \cite{zeng2024tutorial} noted that CKM connects positions, environments, and channel knowledge within a specific region, providing prior channel information for environment-aware communications. Existing studies have used such priors for millimeter-wave beam alignment \cite{wu2021environment}, \cite{dai2024prototyping}, dynamic channel estimation \cite{wu2024environment}, and ISAC sensing \cite{hong2026channel}, \cite{wu2026you}. In these tasks, the system exploits propagation knowledge accumulated in the environment with lower real-time measurement overhead. Therefore, channel knowledge stored in CKM not only corresponds to communication link states, but also encapsulates environmental information relevant to spatial structures and propagation paths. Once introduced into localization, current channel observations can seek corresponding spatial interpretations within the existing map and accomplish location inference.

Long et al. \cite{long2022environment} represented an early effort to introduce CKM into environment-aware wireless localization. They used the LoS information in the CKM as an environment-specific propagation prior to guide greedy anchor selection, achieving localization performance close to that obtained with the optimal anchor set at low computational cost. Wei et al. \cite{wei2025localization} investigated CKM-assisted 3D localization from the perspective of performance analysis. They incorporated multiple types of channel knowledge (e.g., AoA, AoD, and path loss) into location inference, and used virtual APs to describe the geometric relations corresponding to NLoS paths. The results indicate that richer channel knowledge provides more sufficient spatial constraints for location inference, but the gain still depends on map resolution, environmental complexity, and the distinguishability of channel features at different locations.

\begin{figure}[t]
\centerline{\includegraphics[width=0.8\linewidth]{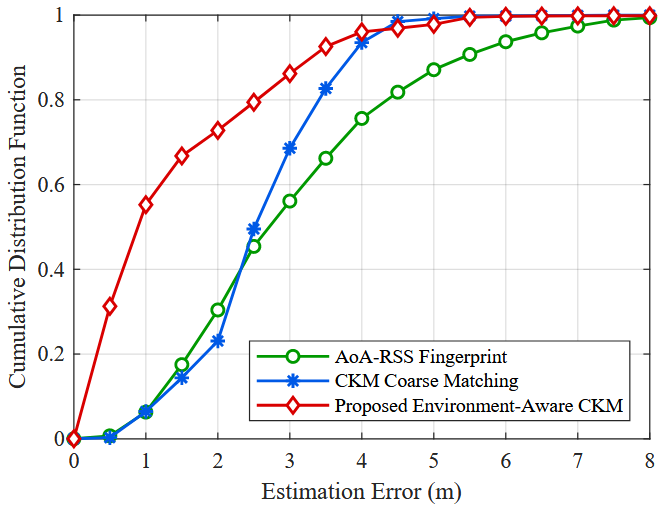}}
\caption{Localization error for environment-aware CKM \cite{hong2026channel}.}
\label{fig:result_hong2026channel}
\end{figure}

CKM not only provides position-related channel priors, but also converts path knowledge accumulated in communication links into environmental priors, providing a robust and extensible solution for sensing and localization in ISAC systems. For urban NLoS localization, \cite{hong2026channel} learned AoA-ToA path features through CKM and mapped each path to candidate scatterers, forming geometric priors for complex environments. In the online phase, observed paths are matched with CKM by similarity to select high-confidence paths, and the main scatterers and user location are jointly estimated through LS. The localization error distributions in Fig.~\ref{fig:result_hong2026channel} show that the proposed method outperforms AoA--RSS fingerprinting and CKM coarse matching in simulations. Similarly, \cite{wu2026you} constructed a channel angle-delay map to connect communication-link modeling with target location inference. These studies demonstrate that CKM can transform multipath information and environmental scattering information, which are difficult to exploit, into spatial geometric constraints and provide new support for localization and sensing in complex environments.

CKM provides a novel perspective for environment-aware localization and shows broad application potential, but this technology is still at an early stage. Existing studies show that, when CKM is applied to localization, current observations are interpreted within existing environmental propagation relations, rather than channel knowledge being merely treated as another type of fingerprint. By linking candidate locations, propagation paths, and environmental scattering structures through channel knowledge, it enables the localization process to accomplish inference through map querying, path matching, node selection, or geometric constraints, reducing reliance on dense online measurements and large-scale labeled fingerprints. CKM has already shown its potential to improve location inference with environmental propagation priors in LoS node selection, 3D localization performance analysis, and NLoS ISAC localization. However, this approach is more suitable for scenarios where environmental structures are stable and infrastructure supports long-term accumulation of channel knowledge. If the map resolution is insufficient, path knowledge is incomplete, or the environment changes, the correspondence between current observations and channel knowledge in the map becomes unstable. At locations with similar multipath patterns or complex scattering structures, inferring positions from channel knowledge may also lead to multiple solutions. Accordingly, constructing more complete CKMs, keeping channel representations in the map accurate and efficiently updated, and designing online query algorithms with low complexity and high accuracy are all problems that must be resolved before practical deployment.

\begin{figure}[t]
\centerline{\includegraphics[width=0.9\linewidth]{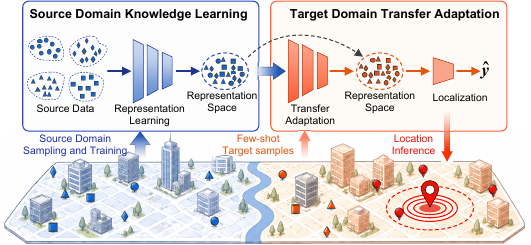}}
\caption{Adaptation of channel representations from source domains to target scenarios, with transferable knowledge adjusted by few target samples for localization.}
\vspace{-2mm}
\label{fig:transfer_adaptation}
\end{figure}

\subsection{Transfer Adaptation-Based Localization}

With advances in model capacity and a deeper understanding of channel measurements, channel representation-based localization methods have achieved high accuracy in many scenarios. However, such accuracy struggles to maintain consistency across different scenarios. Existing models frequently suffer performance degradation or even complete failure due to environment changes. Recollecting scenario data, rebuilding fingerprint databases, or retraining models for each scenario is costly and difficult to satisfy deployment requirements. Thus, a more practical question is how to reuse existing knowledge and adjust model parameters with minimal overhead to achieve rapid adaptation to new scenarios, as depicted in Fig.~\ref{fig:transfer_adaptation}.

\subsubsection{Transfer Learning}
Researchers have attempted to treat feature extractors, representation structures, and parameter knowledge learned from source scenarios as transferable knowledge. Through transfer learning methods such as feature alignment, parameter sharing, and model fine-tuning, the model adapts to new scenarios under the constraint of source domain knowledge. In practice, this can be achieved by directly integrating data from target scenarios during training to minimize distribution discrepancies, or by utilizing target domain data to fine-tune existing networks. In this way, the learned representations establish effective discrimination relations under the new observation distribution.

Fingerprint localization is attractive for practical deployment and can achieve high accuracy, but its performance relies on large-scale fingerprint data \cite{klus2021transfer}, \cite{li2021deep}. Once the scenario changes, it frequently necessitates recollecting data, rebuilding fingerprint databases, or retraining models \cite{stahlke2022transfer}, \cite{liu2017toward}, \cite{zhang2021low}. This makes reducing fingerprint overhead and improving model adaptation highly valuable. Liu et al. \cite{liu2017toward} combined metric learning with metric transfer, importing the distance metric learned in the source domain to the target domain. By constructing a transformation matrix to reshape the target domain sample distribution, the statistical correlations between signal distributions and location labels can be better preserved in the projected feature space. This approach reduces site-survey overhead while maintaining localization accuracy. Kerdjidj et al. \cite{kerdjidj2024exploiting} encoded RSSI into various forms of 2D images and fine-tuned pretrained image networks such as GoogleNet and SqueezeNet, demonstrating that not only can network parameters be transferred, but higher-level representation structures can also be effectively migrated.

\begin{figure}[t]
\centerline{\includegraphics[width=0.75\linewidth]{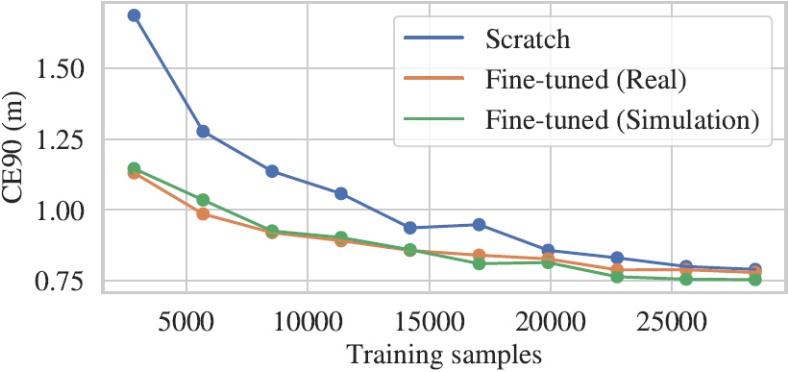}}
\caption{Effect of synthetic and measured CSI pretraining on transfer performance~\cite{stahlke2022transfer}.}
\label{fig:result_stahlke2022transfer}
\end{figure}

Subsequently, attention shifted to what makes transfer effective, how knowledge can be transferred across heterogeneous data, and how representation structures learned in source scenarios can be carried over to target scenarios more smoothly. Li et al. \cite{li2021deep} analyzed the network components for feature extraction and position prediction. They showed that feature extraction layers can be directly transferred to a new environment, whereas the fully connected layers responsible for localization need retraining. For heterogeneous fingerprint spaces caused by changing BS detectability, \cite{li2021transloc} refined the source domain and mapped common and domain-specific BS features into a homogeneous feature space, reducing redundant source knowledge while enabling transfer with very few target samples. Stahlke et al. \cite{stahlke2022transfer} pretrained a CNN-based fingerprinting model on synthetic or measured CSI and fine-tuned it using a small number of target samples. Fig.~\ref{fig:result_stahlke2022transfer} show that pretraining on simulated and measured data yielded comparable localization accuracy after fine-tuning, suggesting that pretrained representations can remain transferable even when the source domain includes simulated data, provided that its propagation structure is sufficiently similar to that of the target scenario. Fontaine et al. \cite{fontaine2023transfer} defined freezing layers and fine-tuning hyperparameters, and used Bayesian optimization to search for the best transfer learning strategy, improving the transfer and reuse of representation knowledge across heterogeneous data. Si et al. \cite{si2023multi} derived the conditions that lead to negative transfer, and used a Bayesian model to select and retain parameter knowledge that promotes positive transfer. This improves knowledge reuse efficiency while addressing the limited personalization of federated learning, negative transfer in transfer learning, and privacy leakage risks.

Reusing established representation knowledge does not represent the ultimate endpoint of transfer adaptation. On the one hand, these approaches attempt to extend learned feature extraction capability, representation structures, and location discrimination experience to new environments, significantly reducing the cost of target scenario data collection, fingerprint reconstruction, and model retraining. However, several works require target domain data to participate in the design of transfer paths, such as shared-feature extraction and cross-domain mapping construction. This makes the adaptation process dependent on a specific target environment and limits cross-scenario applicability. On the other hand, the limited coverage of source scenarios cannot capture all variations and complexity of channel propagation, which restricts model effectiveness and generalization in unseen environments. 

\subsubsection{Meta-learning}
The key idea is to let models ``learn to learn'', treating localization processes in different scenarios as different tasks rather than learning only how to transfer existing knowledge for the sample distribution of a target scenario. By learning from previous tasks, the proposed framework obtains an initialization more suitable for new scenarios, rapidly adjusts channel representations with a few samples in unseen environments, and restores localization capability.

Gao et al. \cite{gao2022metaloc} organized fingerprint localization processes in different scenarios as a set of related tasks, and learned a set of initialization parameters more favorable for novel environments. With these parameters, the model rapidly recovers localization capability after fine-tuning with a few samples. Then, they extended this idea to more general wireless positioning settings, proposing centralized and distributed meta-parameter training strategies for MetaLoc \cite{gao2023metaloc}. Experimental results show that meta-learning learns basic channel representation patterns and alleviates knowledge conflicts induced by excessive environmental discrepancies and overfitting risks caused by limited data. \cite{owfi2023meta} and \cite{jiao2025few} argued that since data distributions across distinct environments are inconsistent, their contributions and importance to the target task should not be treated equally. To this end, \cite{owfi2023meta} introduced task bias, while \cite{jiao2025few} introduced a task weighting mechanism. These designs guide knowledge transfer from training tasks to the target task, improve knowledge transfer efficiency under few-shot conditions, and enable the model to reconstruct channel representations more effectively with limited samples.

Foliadis et al. combined transfer learning with meta-learning. They decomposed representation learning into environment-independent feature extraction and environment-specific feature integration, and embedded knowledge from multiple environments into the former part to strengthen adaptation to unseen environments \cite{foliadis2023multi}, \cite{foliadis2025transfer}. \cite{foliadis2025transfer} proposed a progressive unfreezing strategy to avoid overfitting and catastrophic forgetting during knowledge transfer. The negative log-likelihood loss is used to evaluate uncertainty, effectively reducing the risk of meta-overfitting. This capability of rapid representation reconstruction and adjustment has been extended to more complex deployment settings. \cite{etiabi2024femloc} combines it with federated learning, enabling different terminals to jointly learn a localization model that supports fast personalization without sharing raw data. 

The improvement brought by meta-learning is not limited to reducing the need for labeled samples in the target scenario. Instead, the model pre-learns a base representation form, parameter starting point, or update rule with strong adaptability from multiple related tasks. It emphasizes the accumulation of cross-task experience during meta-training, supporting rapid adjustment of channel representations with a few samples and the rebuilding of position discrimination relations when the model faces unseen environments. However, rapid adaptation capability in meta-learning is not unconditional. Its effectiveness depends heavily on the representativeness of meta-training tasks, the task construction strategy, and the correlation between training tasks and the target environment. The scenarios covered during meta-training determine the nature of the base channel representation the model can learn and how broad a range of propagation variations this representation can handle. If meta-training tasks mainly come from limited or similar environments, the learned initialization and update rules may still be biased toward channel structures in these environments. When the deployment scenario involves more complex blockage, multipath, device differences, or configuration changes, a small number of support samples may be insufficient to recalibrate the representation space, and the rapid adaptation capability diminishes accordingly. In addition, meta-learning typically requires task-set construction and bilevel optimization, while training stability, computational overhead, and task sampling strategies can affect the final transfer effect.

\begin{figure}[t]
\centerline{\includegraphics[width=0.95\linewidth]{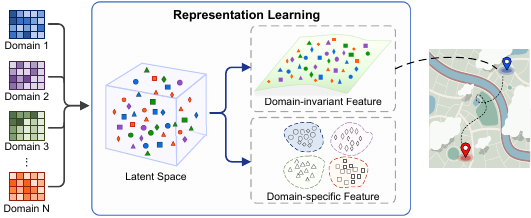}}
\caption{Learning domain-invariant channel representations by separating stable location-related information from domain-specific factors.}
\vspace{-2mm}
\label{fig:invariant_representation}
\end{figure}

\subsection{Invariant Representation-Based Localization}

Practical environments are inherently dynamic. Environmental layout, human activity, device differences, and temporal drift are reflected in channel observations and shift their data distributions, making learned representations difficult to stabilize in out-of-distribution (OOD) scenarios. However, channel observations do not contain only specific environmental information. They also contain relatively stable components governed by underlying propagation laws and related to location. A natural idea is to extract invariant representations from channel measurements that are insensitive to environmental changes while maintaining stable associations with user location. With such representations, the model completes location inference under new environmental or device conditions.

Relevant studies treat channel observations collected under different scenarios, environmental states, dynamic conditions, or device configurations as different domains, where data and feature distributions differ. In terms of representation composition, features that maintain stable associations with location across domains are commonly described as domain-invariant features, whereas domain-specific features denote conditions specific to a domain and components that are difficult to keep stable across domains \cite{xue2025generalization}, \cite{xu2026enhancing}. The former are mainly supported by geometry-related information associated with spatial location, such as propagation distance, angles, and part of the stable multipath structure. The latter are more related to scenario- and system-specific conditions, such as spatial layout, material properties, and device states, and are also affected by human activity, environmental dynamics, and temporal shifts. When a model cannot distinguish these two types of information, representations that appear effective during training may incorrectly treat scenario-specific patterns as location-discriminative evidence, leading to localization performance degradation after deployment conditions change \cite{jiao2025robust}, \cite{chen2022fidora}. For channel representation-based localization, a practical way to improve localization performance and generalization in unknown environments is to learn domain-invariant features that suppress environmental interference and remain stable under domain shifts, as illustrated in Fig.~\ref{fig:invariant_representation}.

\subsubsection{Domain Adaptation}
Domain adaptation has been introduced to constrain feature space distributions through reconstruction, distribution alignment, or adversarial training. These mechanisms suppress unstable perturbations caused by scenario-specific factors and system differences, making channel representations under different conditions more distributionally consistent and more separable by location. 

To address fingerprint inconsistency caused by user differences, Chen et al. \cite{chen2020fido} developed the CSI-based domain adaptation framework FiDo, where a VAE increases fingerprint diversity, and joint constraints from location classification and fingerprint reconstruction constrain the representation space to capture location relevant representations insensitive to user differences. Their follow-up work treated user differences and environmental changes as an unsupervised domain adaptation problem \cite{chen2022fidora}. A clustering assumption is introduced to push target domain samples away from classification boundaries, forcing the feature extractor to learn more common features. For the insufficient expression capability of CSI fingerprints and cross-environment fingerprint inconsistency in dynamic environments, Rao et al. \cite{rao2023mffaloc} reshaped amplitude and phase information into high-resolution fused fingerprint features, strengthening the ability of channel representations to characterize location variations. Maximum mean discrepancy (MMD) constraints guide the constructed two-stream network model to learn invariant representations more stable under environmental changes. 

Considering multipath fading and time-varying fingerprints in long-term deployment, Tian et al. \cite{tian2021wi} treated the initial labeled RSS fingerprint database as the source domain and current unlabeled fingerprints as the target domain. Adversarial generation and feature- and prediction-level cycle-consistency align outdated and current fingerprints in a domain-invariant representation space while preserving location semantics. For signal distribution shifts caused by dynamic changes and device heterogeneity, Lu et al. \cite{lu2025mascloc} embedded RSS, relative fingerprint gradients, and radio maps into a shared multimodal representation space. Domain adversarial learning, together with cosine similarity and Pearson correlation constraints, aligns cross-domain features while preserving the physical consistency of wireless signal variations.

Some studies formulate domain-invariant representation learning as a feature distribution alignment problem. By reducing differences between the source and target domains in global or local statistics, these methods improve the consistency of channel representations. \cite{zhang2024robloc} embeds CSI fingerprints into the Grassmann manifold and aligns global and local source-target distributions, reducing shifts caused by environmental dynamics while preserving location discriminability. For CSI fingerprint drift in dynamic indoor environments, \cite{jiao2025robust} reconstructs calibrated amplitude-phase fingerprints and treats historical data from multiple time scales as source domains. Marginal and conditional distribution alignment is used to transfer stable fingerprint representations to the target domain, which exploits complementary information from different historical environments and reduces the risk of negative transfer when the source domain differs greatly from the target domain.

Domain adaptation methods use target domain observations during training, which enables the model to suppress domain-specific perturbations introduced by environmental, device, and temporal dynamics. Under the joint constraints of source and target domains, the model learns channel representations that remain relatively stable across domains and improve adaptation for environmental changes. However, this approach assumes that target domain data are available. Regardless of whether target domain samples contain location labels or other annotations, they still participate in representation constraints, distribution alignment, or pseudo-label updating during model training. The stable representations are learned from distribution differences within already observed data, yet they may still retain unstable factors relevant to a specific target domain. For practical localization systems, collecting data from target scenarios may be costly or infeasible. Moreover, limited target domain data cannot cover all possible variations. The extracted representations may fail to maintain stable location-related associations under new changes, which limits generalization to unknown scenario variations.

\begin{figure}[t]
\centerline{\includegraphics[width=0.9\linewidth]{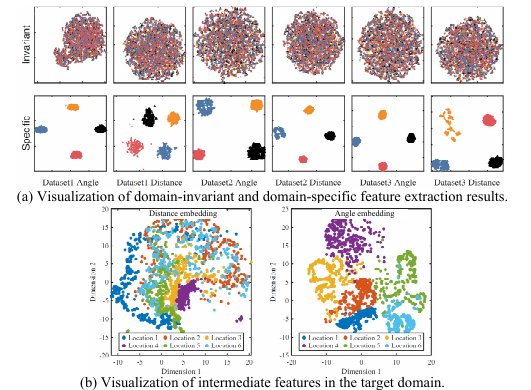}}
\caption{Visualization of disentangled features and target-domain embeddings~\cite{xu2026enhancing}.}
\vspace{-2mm}
\label{fig:featurevisualization}
\end{figure}

\subsubsection{Domain generalization}
Domain generalization offers a more practical perspective for deploying localization systems. It assumes that target scenarios are unseen during training, so models can only use source domain data to learn representations directly transferable to unknown scenarios \cite{zhou2022domain}. Xue et al. \cite{xue2025generalization} learned domain-invariant localization representations by separating them from domain-specific components under mutual information and adversarial constraints. The extracted invariant features maintain localization accuracy in unseen NLoS conditions and physical environments without requiring training data from the new scenarios, highlighting their value for cross-scenario generalization. Xu et al. \cite{xu2026enhancing} combined MMD-based cross-domain alignment with domain-specific feature decoupling to learn fine-grained domain-invariant representations, enabling ranging and angle error mitigation in unseen environments. Feature visualization results in Fig.~\ref{fig:featurevisualization} show that domain-invariant features are mixed across different source domains and exhibit cross-domain consistency, while maintaining a more separable location structure in unseen scenarios. Domain-specific features, by contrast, more explicitly reflect differences among scenarios. These results suggest that feature disentanglement reduces the mixing between stable localization information and scenario-specific perturbations, improving model generalization in unseen environments.

Hu et al. \cite{hu2025dan} argued that existing domain generalization methods may overlook useful domain-specific information when focusing on domain-invariant features. Their domain-level attention network (DAN) uses domain prototypes, attention, and adaptive gating to balance stable localization representations with environment-specific context. Experimental results show that DAN outperforms multiple baselines in unseen environments, indicating that domain-specific information does not necessarily weaken generalization. When properly constrained and fused, it also provides useful complementary information for location inference.

Compared with domain adaptation, domain generalization reduces dependence on target data and is more consistent with rapid deployment in unknown scenarios. By organizing and constraining the representation space, domain generalization suppresses the effect of scenario-specific perturbations on location inference, learns stable and transferable channel representations from complex channel observations, and supports reliable localization and stable generalization. However, no absolute boundary or decision criterion exists between domain-invariant and domain-specific features. Therefore, invariant representation learning should balance cross-domain stability with location discriminability, rather than blindly pursuing stronger cross-domain consistency at the expense of fine-grained clues related to spatial structure. How to characterize the relations between domain-invariant features, domain-specific features, and location information from both theoretical and experimental perspectives, and how to establish more interpretable and verifiable representation constraints, remain challenges that require comprehensive and systematic exploration.

{
\subsection{General Representation-Based Localization}

Prior investigations optimize network structures and model designs to better utilize channel information and learn more expressive channel representations from complex wireless measurements. However, networks trained on scenario-specific data are often limited to specific tasks and data distributions. For multi-scenario and personalized wireless localization, covering different observation forms, tasks, and deployment conditions requires many specialized models, which introduces substantial deployment costs \cite{fontaine2024towards}. At the same time, wireless systems continuously generate large amounts of unlabeled channel observations during pilot transmission, channel estimation, beam training, and link measurements. These data contain rich propagation structures, environmental responses, and spatiotemporal correlations, providing a basis for learning reusable general channel representations.

\subsubsection{General Channel Representations}
By ingeniously designing self-supervised pretraining tasks, models learn propagation structures and channel semantics from massive unlabeled channel observations, forming more general and reusable channel representations \cite{liu2025wifo}, \cite{salihu2024self}. In localization tasks, these general representation models typically serve as channel encoders, mapping channel observations into reusable channel representations. They adapt rapidly to downstream tasks such as geometric parameter estimation, ranging error compensation, and fingerprint localization through task-specific heads or decoders with limited labeled data.

Salihu et al. \cite{salihu2024self} introduced this idea into wireless localization relatively early, using unlabeled data to learn channel features robust to fading variations and system impairments. Stochastic channel augmentations capture macroscopic path loss, shadow fading, and microscopic multipath variations, while the learned representations support few-shot localization. Zheng et al. \cite{zheng2024semi} proposed using differences in channel strength to construct positive sample pairs for contrastive learning and Gaussian Range Encoding to preserve the positions and order of subcarriers within the sequence. They trained the task head using only a few labeled samples, enabling few-shot positioning in unseen areas. To address CSI localization across different frequency bands or bandwidths, Wang et al. \cite{wang2025resilient} developed a mask reconstruction based frequency-scalable localization framework. Multi-band CSI is split into subchannels and converted into sequential data, with masked frequency points reconstructed to support resilient localization with arbitrary numbers of CSI subchannel inputs. 

Pretraining reduces reliance on dense location labels and creates the possibility of reusing channel representations across observation forms, task objectives, and deployment conditions. However, the scope of reuse remains limited by the pretraining data and task design. Therefore, the challenge is to learn propagation structures shared across wireless tasks from channel observations and support adaptation to different wireless systems and downstream tasks.

\begin{figure}[t]
\centerline{\includegraphics[width=0.91\linewidth]{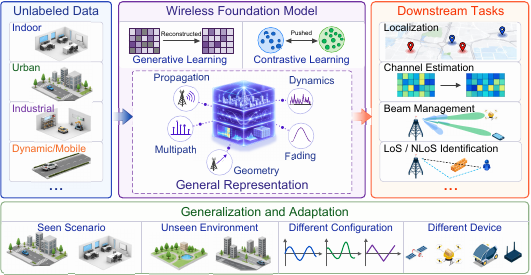}}
\caption{Wireless foundation models learn reusable channel representations from unlabeled observations through self-supervised representation learning, supporting localization and related wireless tasks.}
\vspace{-2mm}
\label{fig:general_representation}
\end{figure}

\subsubsection{Wireless Foundation Models}
For ISAC, tasks such as channel estimation, channel prediction, beam management, environmental sensing, and localization represent different application objectives in the same propagation environment. Although they are implemented in different ways, these tasks are all tied to physical relations embedded in channel observations and governed by electromagnetic propagation, such as geometric information, multipath structure, spatial correlation, and temporal evolution \cite{jiang2025mimo}. Research on general channel representations is increasingly turning to foundation models (Fig.~\ref{fig:general_representation}) pretrained on large, diverse datasets \cite{yazdnian2026multi}. These models can be adapted to and reused across communication, sensing, and localization tasks \cite{guler2025multi}, while retaining transferability and robustness under fading variations, system impairments, environmental perturbations, and changes in observation configurations \cite{khan2026comprehensive}.

\textbf{Single-Task Pretraining}. Ott et al. \cite{ott2024radio} used unlabeled CIR data and pretrained a transformer by randomly masking, dropping, and reconstructing CIR samples to learn spatio-temporal propagation patterns. With limited labeled data, the pretrained model achieves fingerprint localization accuracy comparable to supervised pretraining. Cheraghinia et al. \cite{cheraghinia2025foundation} considered adaptation problems caused by heterogeneous data types, sampling rates, and environmental changes in wireless technology recognition and localization. Their foundation model takes I/Q signals and CIRs as inputs and adapts to wireless technology recognition, LoS/NLoS identification, and ranging error compensation through lightweight fine-tuning. The results show broad task adaptability, but performance degradation and negative transfer may occur in unseen localization-related scenarios. Jiang et al. \cite{jiang2025mimo} aligned CIR-CSI positive pairs in a shared embedding space to jointly exploit time-domain multipath structures and frequency-domain channel responses. This consistency improves representation reuse and task adaptation across positioning, beam management, and channel identification. Compared with supervised learning, this approach reduces the mean error distance in positioning by approximately 22\%.

\textbf{Hybrid Pretraining}. Channel representations learned from a single task tend to reflect only a certain aspect of channel characteristics. To overcome this limitation, pretraining has been extended to hybrid pretext tasks. With constraints such as spatio-temporal information recovery and cross-modal consistency, the learned representations become more expressive and complete. Guler et al. \cite{guler2025multi} combined CSI-based masked reconstruction with contrastive learning, so that the learned representations capture both structural and discriminative channel information. The representations support channel estimation, beam selection, and LoS identification, and show robustness under cross-frequency, cross-environment, and low-SNR conditions. Pan et al. \cite{pan2025large} analyzed self-supervised tasks from the information bottleneck perspective and jointly optimized spatial-frequency masked channel modeling, domain-transform consistency, and location-invariant contrastive learning. The model maintains stronger performance under label-limited fine-tuning and unseen configurations.

\textbf{Multimodal Observation Inputs}. To learn unified representations across wireless modalities, Aboulfotouh et al. \cite{aboulfotouh20256g} represented spectrograms, CSI, and I/Q signals as image-like inputs to a shared backbone and developed the WavesFM foundation model. In later work, they noted that the most informative modality varies with the task and operating conditions, and no single modality is universally optimal. They used modality-specific input embeddings to map I/Q streams and image-like wireless observations (e.g. CSI and spectrograms) into a common token space. A shared encoder learns modality-agnostic wireless representations from these token sequences, extending the range of observations and tasks supported by the same backbone \cite{aboulfotouh2026multimodal}.

\textbf{Physical Environment Information Fusion}. \cite{jiao2025addressing} and \cite{yazdnian2026multi} extend general channel representation learning from wireless observations to cross-modal relations between channel observations and physical environments. Both studies regard channel measurements, spatial geometry, and environmental structures as coupled through electromagnetic propagation, using this relation to guide the learning of shared representations between wireless observations and explicit environmental information. \cite{yazdnian2026multi} incorporated CSI, scene layout, and user locations into a masked reconstruction framework, where a global aggregation representation summarizes cross-modal physical information and reconstructs the spatial spectrum. This links the learned representation to propagation directions and energy distributions, supporting transfer to position estimation, channel estimation, and multi-antenna optimization under limited labeled data and unseen scenarios. \cite{jiao2025addressing} aligned explicit scattering environment descriptions and CSI in a shared embedding space, with reconstruction and cross-modal sharing modules used to preserve propagation path information. In downstream adaptation, the frozen pretrained model supports direct zero-shot inference and pluggable lightweight fine-tuning for unseen scenario tasks. By linking channel states with explicit environmental semantics, these works introduce more interpretable physical constraints and improve the cross-task reuse and cross-scenario adaptation of channel representations.

\textbf{LLM Adaptation}. Rather than training wireless-specific models from scratch, recent studies adapt channel observations to pretrained large models through input mapping and lightweight tuning. Ozeki et al. \cite{ozeki2025cellular} reused pretrained GPT-2 as a sequence feature extractor by mapping RSS sequences and BS IDs into its latent space, and used label-aware contrastive learning to make feature distances consistent with physical distances. This approach partially avoids the heavy cost of network architecture design and training from scratch. It also shows that sequence modeling and structural representation capabilities formed in pretrained large models transfer to wireless channel data and produce channel representations for downstream tasks such as channel prediction, channel estimation, and wireless localization.

\textbf{Downstream Tasks}. Foundation models aim to learn general, reusable representations for multiple downstream tasks and have shown promise in wireless communication, sensing, and localization. Downstream tasks include:
\begin{itemize}
    \item Sensing and Localization: Wireless observations are used as carriers of physical information to infer spatial, behavioral, and environmental states. Typical tasks include localization \cite{pan2025large}, geometric parameter estimation \cite{pan2025large}, propagation state identification \cite{cheraghinia2025foundation}, activity sensing \cite{aboulfotouh20256g}, and environment reconstruction \cite{zeng2024tutorial}.

    \item Communication: These tasks perform signal processing at the transmitter and receiver, using wireless signals as carriers of communication information, and include semantic communication \cite{qiao2025todma}, signal recognition \cite{cheraghinia2025foundation}, denoising \cite{luo2025emind}, and interference detection and classification \cite{aboulfotouh2026multimodal}.

    \item Channel Processing: The aim is to acquire, predict, and efficiently convey channel information that reflects wireless propagation characteristics. Existing studies address channel modeling \cite{han2026wico}, channel estimation \cite{yazdnian2026multi}, channel recovery \cite{guler2025multi}, channel prediction \cite{liu2025wifo}, CSI feedback \cite{jiao2025addressing}, and path loss estimation \cite{liu2025llm4wm}.

    \item Beam Management: It determines directional transmission and reception strategies to maintain link quality and communication efficiency, e.g., beam selection \cite{guler2025multi}, beam prediction \cite{alikhani2025lwm}, and hybrid precoding \cite{yazdnian2026multi}.
\end{itemize}

Overall, the foundation model no longer treats localization as a closed problem requiring independent training. It learns a reusable representation foundation from large-scale channel observations and adapts it to specific tasks with limited labeled samples and lightweight task heads. More importantly, it redirects representation learning from scenario-specific or task-specific feature extraction toward base representations for a class of wireless propagation problems, providing a new direction for low-cost deployment of high-accuracy localization. However, the effectiveness of foundation models depends on pretraining data and adaptation mechanisms. Even with a foundation model, a small amount of labeled data from the target scenario or task is still needed for fine-tuning. Existing studies mainly verify model capability on simulated data, limited scenarios, or specific downstream tasks, and their usability in real complex environments requires systematic examination. The storage, computation, and training costs introduced by large-scale pretraining should also not be ignored. Future studies should go beyond model architecture design, input modality selection, and pretraining data expansion, and provide stronger evidence that base representations truly capture key propagation information such as channel states and multipath structures, and that such information stably supports location inference under real environmental changes.}

\begin{table*}[t]
\caption{Comparison of channel representation-based localization methods.}
\label{tab:representation_based_localization}
\centering
\scriptsize
\renewcommand{\arraystretch}{1.2}
\setlength{\tabcolsep}{2.2pt}
\begin{lrbox}{\localizationtablebox}
\begin{tabular}{@{}l@{\hspace{4pt}}l@{\hspace{2pt}}|c|c|c|c|>{\centering\arraybackslash}p{0.11\textwidth}|c|p{0.165\textwidth}}
\hline
\multicolumn{2}{c|}{\multirow{2}{*}{\textbf{Method}}} &
\multicolumn{1}{c|}{\multirow{2}{*}{\textbf{Accuracy}}} &
\multicolumn{1}{c|}{\multirow{2}{*}{\shortstack[c]{\textbf{Source}\\\textbf{Labeled Data}}}} &
\multicolumn{3}{c|}{\textbf{Generalization}} &
\multicolumn{1}{c|}{\multirow{2}{*}{\textbf{Cost}}} &
\multicolumn{1}{c}{\multirow{2}{*}{\textbf{References}}} \\
\cline{5-7}
\multicolumn{2}{c|}{} & \multicolumn{1}{c|}{} & \multicolumn{1}{c|}{} &
\textbf{Capability} & \textbf{Cross-domain} & \textbf{Target Data} & \multicolumn{1}{c|}{} & \multicolumn{1}{c}{} \\
\hline\hline
\multirow{3}{*}{\textbf{Classical Representation}} & Handcrafted feature & \medstar & \medsq & \medtri & $\times$ & -- & \lowcirc & \cite{guvencc2007nlos} \cite{marano2010nlos} \cite{wymeersch2012machine} \cite{ferreira2021feature} \\
 & Fingerprint & \medstar & \highsq & \lowtri & $\times$ & -- & \highcirc & \cite{bahl2000radar} \cite{xiao2012fifs} \cite{chen2020aoa} \\
 & Deep learning & \highstar & \highsq & \lowtri & $\times$ & -- & \highcirc & \cite{bregar2018improving} \cite{kim2022uwb} \cite{yang2022robust} \cite{zhang2023csi} \\
\hline
\multirow{2}{*}{\textbf{Spatial Structure}} & Channel charting  & \lowstar & \lowsq & \lowtri & $\times$ & -- & \lowcirc & \cite{pihlajasalo2020absolute} \cite{stahlke2023indoor} \cite{taner2025channelcharting} \cite{stephan2025three} \cite{poeggel2025passive} \\
 & Channel knowledge map & \medstar & \medsq & \medtri & $\times$ & -- & \highcirc & \cite{long2022environment} \cite{wei2025localization} \cite{hong2026channel} \cite{wu2026you} \\
\hline
\multirow{2}{*}{\textbf{Transfer Adaptation}} & Transfer learning & \highstar & \highsq & \medtri & \checkmark & \targetdata{\medsq}{Fine-tuning} & \medcirc & \cite{liu2017toward} \cite{li2021transloc} \cite{stahlke2022transfer} \cite{guo2023fedpos} \cite{si2023multi} \\
 & Meta-learning & \highstar & \highsq & \medtri & \checkmark & \targetdata{\lowsq}{Fine-tuning} & \highcirc & \cite{gao2023metaloc} \cite{owfi2023meta} \cite{jiao2025few} \cite{foliadis2025transfer} \cite{etiabi2024femloc} \\
\hline
\multirow{2}{*}{\textbf{Invariant Representation}} & Domain adaptation & \highstar & \highsq & \medtri & \checkmark & \targetdata{\medsq}{Alignment} & \medcirc & \cite{chen2020fido} \cite{chen2022fidora} \cite{rao2023mffaloc} \cite{tian2021wi} \cite{zhang2024robloc} \\
 & Domain generalization & \highstar & \highsq & \hightri & \checkmark & $\times$ & \highcirc & \cite{xue2025generalization} \cite{xu2026enhancing} \cite{hu2025dan} \\
\hline
\multirow{1}{*}{\textbf{General Representation}} & Self-supervised learning & \highstar & $\times$ & \medtri & \checkmark & \targetdata{\lowsq}{Fine-tuning} & \highcirc & \cite{salihu2024self} \cite{ott2024radio} \cite{guler2025multi} \cite{pan2025large} \cite{jiao2025addressing} \\
\hline
\end{tabular}
\end{lrbox}
\usebox{\localizationtablebox}
\par\vspace{2pt}
\parbox{\wd\localizationtablebox}{\raggedright\scriptsize\textbf{\textit{Note:}} One, three, and five symbols indicate low, medium, and high levels, respectively; \checkmark\ and $\times$ indicate yes and no.}
\end{table*}

\begin{figure}[t]
\centerline{\includegraphics[width=0.9\linewidth]{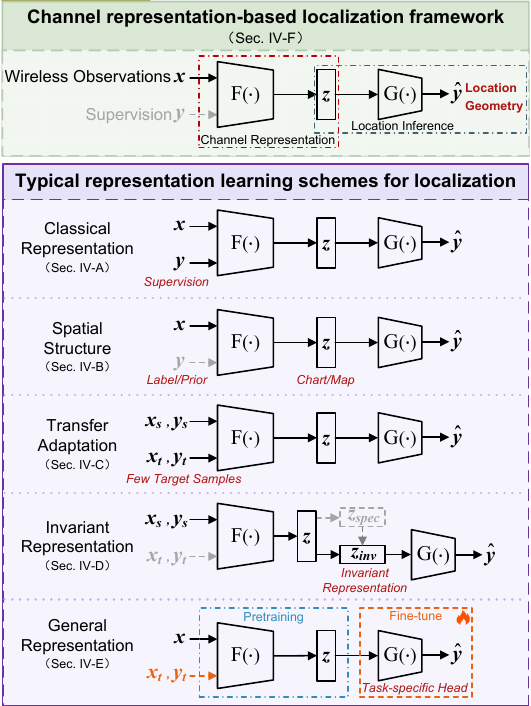}}
\caption{Channel representation-based localization framework and typical representation learning schemes for localization. Gray elements denote optional inputs or branches that may be absent in specific implementations.}
\vspace{-2mm}
\label{fig:localization_framework}
\end{figure}

{
\subsection{Channel Representation-Based Localization Framework}
\subsubsection{Framework Formulation}

From an information processing perspective, learning-driven localization should not be interpreted merely as an end-to-end mapping from wireless observations to locations.  Instead, it typically requires extracting compact channel representations from raw channel measurements before performing location inference. The localization performance of a model depends not only on the final inference module, but also on the location information encapsulated within the representation and the quality of that information. Grounded in this insight, this paper models the localization process as a unified ``wireless observation--channel representation--location inference'' framework.

The framework in Fig.~\ref{fig:localization_framework} consists of two cascaded modules, namely the channel representation (feature extraction) module $F(\cdot)$ and the location inference module $G(\cdot)$. Specifically, $F(\cdot)$ encodes the channel observation $\bm{x}$ into the channel representation $\bm{z}$, formulated as $\bm{z}=F(\bm{x})$. The module $G(\cdot)$ uses $\bm{z}$ to estimate the location or geometric parameters, i.e., $\hat{\bm{y}}=G(\bm{z})$. The output $\hat{\bm{y}}$ may denote the absolute coordinates, relative position, distance, angle, delay, trajectory, or other localization-related quantities.

From this perspective, $F(\cdot)$ dictates the quality of the representation, while $G(\cdot)$ determines its utilization. Functionally, $F(\cdot)$ is not only a feature extractor, but also a module for organizing, optimizing, and constraining representations. $G(\cdot)$ maps the representation $\bm{z}$ to localization-related outputs via diverse mechanisms such as matching, classification, regression, geometric computation, graph reasoning, or sequence tracking. While the complexity of $G(\cdot)$ varies across localization tasks, its performance is constrained by the information quality encapsulated within representations. Fundamentally, the key problem in channel representation-based localization is to construct effective and high-quality representations, rather than merely infer locations from observation.

The framework shifts the focus of model design and analysis from localization error to the formation, organization, and information quality of channel representations, emphasizing more stable and expressive representations for accurate, robust, and generalizable location inference. Fig.~\ref{fig:localization_framework} illustrates different realizations of this framework, and Table~\ref{tab:representation_based_localization} summarizes their characteristics.

The effectiveness of a representation for localization depends on whether it retains sufficiently discriminative and stable information for the task. Physically meaningful features can support error compensation or provide geometric constraints for localization at low cost. Geometric relations among samples or prior knowledge of the spatial environment can provide physical constraints for representation learning, reducing reliance on dense location labels but introducing costs for structure construction and coordinate calibration. Cross-domain generalization requires channel representations to retain location discriminability in the target domain. This can be achieved through stable representations learned from source data or by adapting existing representations using target data. Stable representations must preserve distinctions between locations while suppressing scenario-specific disturbances. However, techniques such as information disentanglement, cross-domain alignment, and large-scale pretraining increase model design complexity and training costs. And in general, target-domain data are difficult to obtain in practice, and limited samples may not capture all propagation variations in the target scenario. Therefore, representation design should balance location discriminability, stability across scenarios, and construction cost given the localization requirements, data availability, and deployment conditions.}

{
\subsubsection{Information Bottleneck Analysis}

The channel representation-based localization framework presents the flow of location-related information during localization. Information bottleneck (IB) theory \cite{tishby2000information} can be used to analyze the information constraints among wireless observations, channel representations, and location inference.

Let $X$, $Z$, $Y$, and $\hat{Y}$ denote the wireless channel observation, channel representation, true location, and estimate, respectively. The objective is to learn representation $Z$ from the observation $X$ associated with location $Y$ and accurately estimate the location or geometric parameters $\hat{Y}$. From an information transfer perspective, this process can be represented by the Markov chain $Y \rightarrow X\rightarrow Z \rightarrow \hat{Y}$. By the data processing inequality \cite{cover1999elements}, $I(Y;X)\geq I(Y;Z)\geq I(Y;\hat{Y})$. Location-related information lost at any stage cannot be recovered or used downstream. Thus, the information contained in the representation forms the basis for location inference.

The ideal channel representation should satisfy task sufficiency and information minimality \cite{shamir2010learning}. Task sufficiency requires $Z$ to retain all information in $X$ relevant to inferring $Y$, so that $X$ provides no additional location-related information once $Z$ is known. Information minimality requires discarding observation components that do not contribute to location inference. In practice, finite samples, limited model capacity, and optimization constraints make both requirements difficult to satisfy simultaneously. The IB objective $\mathcal{L}_{\mathrm{IB}} = I(X;Z) - \beta I(Y;Z)$ balances task sufficiency and information minimality \cite{tishby2000information}. Here $I(X;Z)$ and $I(Y;Z)$ measure the observation information and location-related information encoded in the representation, respectively, and $\beta>0$ is a weight.

Localization algorithm design should jointly consider channel representation and location inference. Representation construction should retain channel information that distinguishes locations while reducing task-irrelevant observation components. Location inference should exploit this information effectively to improve performance. The analysis specifies theoretical information requirements for ideal channel representations, but translating these requirements into computable and comparable metrics remains challenging. Existing studies estimate mutual information using noise injection, discretization, or variational bounds, but these estimates are subject to bias due to noise assumptions, bin widths, variational approximations, and finite sample sizes. Therefore, methods for quantifying channel representation quality require further refinement.}

\section{Challenges and Future Directions}
\label{sec:challenges}

Channel representation-based localization extends learning-driven localization beyond end-to-end position fitting. It provides a way to interpret localization evidence from wireless propagation and spatial structure, creating new possibilities for high-accuracy localization. However, translating these methods from controlled experiments into real deployment still faces a series of challenges that warrant further study.

\subsection{Data Collection}

\subsubsection{Collection Cost}

Accurate position labels require additional equipment, such as real-time kinematic (RTK) receivers, laser trackers, or light detection and ranging (LiDAR) systems. This increases collection cost and makes large-scale, long-term, multi-scenario labeling difficult. For scenario information, satellite maps, architectural drawings, building information modeling models, or 3D point clouds serve to construct environmental representations, but this information also requires additional collection, registration, and maintenance. For channel observations, metadata such as frequency band, bandwidth, antenna configuration, transceiver devices, coordinate system, and environmental state must be recorded synchronously. Low-quality or abnormal samples also need to be removed through anomaly detection, channel quality assessment, data cleaning, and phase calibration.

\subsubsection{Limited Data Coverage}

The more challenging issue is that wireless signals continuously interact with the environment during propagation. Channel observations are governed by physical laws and affected by both quasi-static factors (e.g., spatial layout, material properties, and blockage structures) and dynamic factors (e.g., vehicle motion, human activity, object movement, and transceiver state changes), making them highly sensitive, diverse, and complex. The collected data cover only part of the propagation states at a specific time and in a specific scenario, making it difficult to fully describe all variations in real environments. With aging environments, device replacements, or altered system configurations, existing data gradually become outdated and require continuous updating and maintenance. Meanwhile, wireless systems continuously generate massive channel data during pilot transmission, channel estimation, beam training, and link measurement. Unfortunately, stringent privacy protections, restrictive access permissions, and closed system architectures hinder the collection and processing of these valuable data. With the development of massive MIMO, millimeter-wave/terahertz systems, wideband OFDM, reconfigurable intelligent surfaces (RIS), and cell-free networks, the dimensionality of channel data escalates drastically, which significantly raises the cost of collection, transmission, storage, and processing. Consequently, these factors lead to a limited number of real-world public datasets and insufficient scenario coverage.

\subsubsection{Dataset Standardization}

A more pressing challenge is that there is still no standardized procedure or technical guideline for dataset collection, processing, and quality control. Due to inconsistencies in coordinate systems, device configurations, sampling methods, label accuracy, and environmental descriptions, different datasets are difficult to merge, and models trained on one dataset are difficult to reuse on another.

\subsubsection{Simulation-to-Reality Gap}

Synthetic data provide another way to alleviate the scarcity of real data. Ray tracing, stochastic channel models, digital twins, and electromagnetic simulation generate large-scale channel samples under controllable conditions, while explicitly providing information such as positions, paths, and environmental geometry. These data are useful for model pretraining and analyzing representation learning mechanisms. However, the quality of synthetic data depends on propagation models, scene modeling accuracy, and simulation parameter settings. Synthetic data characterize idealized propagation structures, but they still cannot fully reproduce dynamic changes, hardware impairments, synchronization errors, measurement noise, and human activities in the real world. Models trained only on synthetic data require real measurement data for calibration or fine-tuning. Narrowing the gap between simulated and real data remains a challenge for data-driven localization.

\subsection{Heterogeneous Sources}

\subsubsection{Complementarity}

Constructing location-related representations around channel observations does not mean that a localization system relies only on a single form of wireless measurement. Different wireless observation forms describe propagation characteristics from perspectives such as signal strength, time-frequency response, and geometric parameters. Measurements from diverse wireless technologies, such as UWB, WiFi, Bluetooth, and cellular systems, can work alongside inertial, visual, digital maps, and scenario priors to provide position constraints. Fusing multisource heterogeneous information compensates for the limitations of a single channel observation, gives channel representations more complete propagation structures, spatial constraints, and environmental semantics, and improves the expression of stable location-related information. Fusion effectiveness is the basic prerequisite for utilizing heterogeneous information. Over-fusion may introduce modality noise irrelevant to localization, while insufficient fusion increases data collection and model training costs without fully exploiting the complementarity among different observations.

\subsubsection{Cross-Source Alignment}

However, the existence form and granularity of location information vary across measurements. Wireless signals imply spatial information through amplitude, phase, delay, and angle. Inertial measurements reflect short-term motion variations, and visual or map information describes spatial layouts and environmental structure more thoroughly. Even within wireless observations, signals from different technologies and frequency bands have distinct propagation mechanisms, coverage ranges, and measurement accuracy. For example, UWB provides high temporal resolution, WiFi and cellular systems provide CSI or link measurements, and low-power signals such as Bluetooth are more suitable for low-cost deployment. Low-frequency signals tend to have stronger diffraction capability, whereas high-frequency signals more readily provide fine spatio-temporal resolution. Different information sources differ significantly in resolution, spatial reference, and error characteristics. Consequently, time synchronization, spatial registration, and scale unification are required in many cases, and issues such as modality missingness, blockage, drift, and time-varying sensor reliability must also be considered.

\subsubsection{Fusion-Oriented Source Evaluation}

The main challenge of multisource fusion is not merely feeding different observations into the same model, but determining the role boundary of different information sources in channel representation learning. Wireless observations from different technologies and frequency bands can enrich channel representations from aspects such as propagation strength, multipath structure, and geometric information. External information also enhances representations as geometric priors or cross-modal consistency constraints. At the same time, these sources may introduce new unstable factors due to differences in noise characteristics, scales, and reliability. Mechanisms to quantify the contribution, reliability, and complementarity of different information sources are lacking. Analytical methods that explain how different observations affect channel representation structures and location inference results are also missing. A key problem in multisource heterogeneous fusion is how to make multiple wireless observations and external information, such as vision, inertial measurements, maps, and environmental semantics, effective complements to channel representations rather than sources of additional distribution shifts or redundant dependence. Designing scalable models that can adapt to the different characteristics of information sources while remaining robust to source absence is an additional challenge.

\subsection{Representation Mechanism}
\subsubsection{Modeling with Propagation Knowledge}

Feature extraction and representation learning have achieved remarkable success in computer vision and natural language processing, providing a wealth of model architectures and training strategies that can be leveraged for channel representation-based localization. However, wireless signals are not simple substitutes for images or text. Image and text data also have complex structures, but mature modeling approaches have been developed. In contrast, wireless observations are influenced by electromagnetic propagation laws, system configurations, and environmental geometry, and their structural regularities differ from those of images and text. Existing studies have introduced machine learning methods, deep network models, and pretraining strategies into channel feature extraction and representation learning, achieving favorable results in localization tasks. Yet the relationship between these methods and wireless propagation mechanisms remains insufficiently understood. This limitation appears not only in model structures, but also in training objectives and data processing. For instance, data augmentation may change channel amplitude-phase relations and spatial correlations, masked reconstruction may disturb delay distributions and multipath structures, and positive and negative sample construction in contrastive learning may not align with spatial correlations and channel similarities in actual propagation. Furthermore, many model structures and loss functions are derived from general artificial intelligence (AI) experience rather than from physical knowledge such as channel models, localization identifiability, or propagation geometric relations.

\subsubsection{Representation Interpretability}

Although wireless localization performance has steadily improved through exploration of channel representations, this advancement does not mean that the underlying representation mechanism is already clear. AI-driven data methods have powerful feature extraction capability, but they tend to lack interpretability and rarely reveal the intrinsic connections between the environment and wireless channel. Existing studies mostly demonstrate that certain feature extraction methodologies can improve localization performance, without providing a mechanistic analysis of the underlying reasons for their validity or applicability. High-dimensional channel observations simultaneously contain geometry-related propagation information associated with location, response features introduced by devices and system configurations, and perturbation factors caused by environmental changes. Consequently, the extracted features may couple location-related information, system-specific information, and scenario-specific information. Qualitative and quantitative analyses are lacking for the information contained in learned channel representations, the components that truly correspond to geometric information, and the performance boundaries supported by these features. Systematic characterization is also needed for the effects of observation conditions (e.g., array aperture, bandwidth, and SNR) on representation quality and performance limits, as well as their propagation to geometric information estimation and location inference.

\subsubsection{Theoretical Characterization}

Answering these questions requires revisiting channel representation through information preservation, parameter estimability, and representation-space structure, rather than relying solely on empirical comparisons of localization error. The information bottleneck \cite{pan2025large}, \cite{tishby2000information} and mutual information analysis \cite{xue2025generalization} serve as potential tools to discuss whether representations retain location-related information while compressing raw channel observations, and whether they mitigate domain-specific factors such as scenario, device, and time. Fisher information and the CRLB \cite{shen2010fundamental}, \cite{li2019crlb}, \cite{han2016performance} can characterize the amount of information available in channel observations for delay, angle, range, and position estimation from a parameter estimation perspective, providing references for understanding localization error lower bounds and representation performance limits. Furthermore, manifold learning and topological analysis \cite{studer2018channel}, \cite{ferrand2021triplet}, \cite{stahlke2023indoor} examine whether the representation space preserves neighborhood relations in physical space. Causal representation learning and invariant risk minimization \cite{peters2016causal}, \cite{arjovsky2019invariant}, \cite{scholkopf2021toward} provide possible insights for distinguishing stable location factors from scenario-specific spurious correlations. Although these tools have not yet formed a unified theoretical framework tailored for channel representation-based localization, they offer promising directions for explaining why representations work, when they fail, and whether they can generalize.

\subsection{Generalization Capability}

\subsubsection{Distribution Shift}

Like other learning-driven methods, channel representation-based localization inevitably faces generalization challenges caused by data distribution shifts. Distribution shifts in wireless localization arise from multiple levels. Environmental layout, human activity, object movement, and temporal aging change the multipath structure. Device differences, system configurations, and sampling strategies change the amplitude, phase, and statistical properties of channel observations. These shifts stem from the coupling between wireless propagation and environmental or system conditions, and are difficult to eliminate through general data processing operations such as data augmentation or normalization. Models trained under specific scenarios and conditions may suffer noticeable performance degradation, or even complete failure, when deployment conditions change or the model is transferred to other scenarios. Ensuring model universality and stability under different scenarios and conditions is a critical challenge that must be solved.

\subsubsection{Geometry-Related Stability}

Channel representation provides an explanation from the perspective of information composition. Channel information comprises spatial geometry-related information and environmental perturbation information. If models treat scenario-specific multipath patterns or system responses as evidence for position estimation during training, these features may no longer remain stably associated with spatial geometry when the environment and deployment conditions change, and localization performance will degrade accordingly. Transfer learning, meta-learning, domain adaptation, domain generalization, and foundation models alleviate this problem. However, they cannot fully solve it due to limitations such as the representativeness of source-domain data and the availability of target domain data. CC and CKM, which place greater emphasis on spatial structure, have not completely avoided the generalization problem either. CC reduces dependence on absolute positions, but systematic methods are lacking for keeping the learned relative structure consistent across scenarios and mapping it stably to real coordinate systems. CKM is more suitable for quasi-static environments, while map maintenance, knowledge updating, and dynamic adaptation remain difficult when facing human activity, blockage changes, and device-configuration updates.

\subsubsection{Generalization Characterization}

More importantly, existing studies mostly rely on empirical analysis to show that certain channel components or representation components support model generalization and universality. Substantive evidence is still lacking for why these components remain stable and how they support location inference. Under current evaluation methods, localization errors in different scenarios only indicate whether performance degradation occurs, but they do not explain the main source of the degradation among channel observation distributions, representation space structures, or inference modules. Due to the lack of characterization of the ``channel observation--representation space--localization error'' transfer chain, existing methods tend to verify generalization only through empirical experiments. Consequently, it is difficult to determine the presence of information that maintains generalization stability within the representation, as well as the specific conditions for such information to remain stable or become invalid. The generalization problem must be pushed beyond error-oriented evaluation toward a mechanistic analysis of the formation, variation, and transfer of channel representations. A generalization analysis framework for channel representations is required to characterize the effects of distribution shifts on representation spaces and localization errors. Such a framework serves to guide the design of more targeted representation constraints, physical consistency modeling, and low-cost adaptation methods.

\subsection{Standardized Evaluation}

\subsubsection{Benchmarks}

Currently, the wireless field lacks a unified evaluation foundation comparable to large standard databases in computer vision or standard corpora in natural language processing. There is no standardized data and evaluation system established for wide adoption that covers multiple scenarios, distinct devices, and diverse system conditions. Most studies rely on self-collected datasets or a few public datasets, which correspond to different scenarios, devices, and system configurations and differ significantly in sampling density, label accuracy, and preprocessing. Accordingly, similar localization accuracy reported by different models does not indicate comparable representation quality or adaptation capability. Inconsistencies in data processing, dataset partitioning, and evaluation methods further weaken the comparability of results. Preprocessing steps such as phase calibration, normalization, and filtering change representation distributions and model performance, yet many works do not fully report these details. Localization results also tend to rely on dense sampling, accurate labels, complex pretraining, or target-scenario fine-tuning, while data collection volume, labeling cost, inference time, and computational overhead are rarely discussed.

\subsubsection{Representation Metrics}

Existing evaluation metrics mainly focus on final task outputs, such as localization error, ranging or angle error, and classification accuracy. While final task metrics are necessary for measuring system performance, they do not directly demonstrate the effectiveness of representation. The localization performance of the model may come from the quality of channel representations, but it may also come from a more complex localization head, denser sampling, more sufficient data augmentation, or a more favorable dataset split. Channel representations are abstract and difficult to interpret or verify, unlike localization errors which can be directly observed and compared. Consequently, more concrete analyses are required to determine the specific information truly preserved and to evaluate the quality and efficiency of the representations. Prior works have visualized feature distributions and their relationships with geometric information via dimensionality reduction \cite{xu2026enhancing}, \cite{jiao2025addressing}, \cite{yazdnian2026multi}. However, low-dimensional projection may change sample distances and local structures, and the distributions shown in 2D/3D space cannot fully represent the real geometric structure in the high-dimensional representation space. Such visualization can only serve as auxiliary evidence for representation rationality. 

\subsubsection{Evaluation Protocols}

Therefore, future evaluation frameworks must shift from isolated error and precision assessments toward more comprehensive representation validation. It is essential to investigate the representation space itself and assess the information metrics, expressive capability, and transferability of channel representations by numerical or visual methods. In addition to fully documenting data collection, processing procedures, and experimental settings, studies should report the amount of data required for model adaptation, computational overhead, and deployment cost, and introduce metrics that reflect representation structure, effectiveness, and transferability. Only by discussing final localization performance alongside representation quality, generalization conditions, and deployment costs within a unified framework can the actual value of different methodologies be accurately determined.

\subsection{Deployability}

\subsubsection{Maintenance Cost}

Deployability primarily comes from model maintenance and transfer. A deployable model should achieve low error in a single scenario and support cross-scenario reuse, low-cost updating, and rapid adaptation. Most learning-driven localization methods achieve high accuracy on a specific dataset. Once deployed in a new scenario, device, or system configuration, they may require new data collection, model retraining, or adaptation. This makes the real deployment cost of a model much higher than the performance advantage reflected only by localization error in experiments.

\subsubsection{Resource Constraints}

Learning-based models require substantial computation, memory, and energy during training and inference. Therefore, model design should consider not only localization accuracy, but also deployability under real device and system conditions. Computing capabilities and resources vary significantly across different terminals. For ultra-low-power devices (e.g., IoT sensors and smart wearables), collecting data, fine-tuning models, or running complex neural networks is almost impossible, and extremely compact models are the only practical choice for local inference. Edge devices (e.g., smartphones, unmanned aerial vehicles, and mobile robots) have certain computing capability, but they mainly support lightweight fine-tuning, incremental updating, or low-frequency model adaptation. Inference latency, power consumption, storage footprint, and communication overhead must also be considered. For time-sensitive scenarios (e.g., autonomous driving, industrial robots, and emergency rescue), model inference time and system response delay directly affect the usability of localization methods. Incorporating computational resources, energy constraints, real-time requirements, and cross-scenario maintenance costs into model design to balance accuracy and deployability is a challenging task.

\subsection{Trustworthiness, Security, and Privacy Protection}

\subsubsection{Uncertainty Estimation}

In high-risk scenarios such as autonomous driving, emergency rescue, and personnel safety monitoring, it is necessary not only to report position estimates, but also to evaluate confidence, abnormal states, and failure risks. Learning-driven models typically extract representations in a black-box manner. When the environment changes, observation quality decreases, or input samples deviate from the training distribution, the model may still output seemingly certain localization results, while lacking reliable estimates of error and uncertainty. Introducing uncertainty modeling, anomaly detection, and confidence assessment into channel representations is another practical problem.

\subsubsection{Security and Privacy Protection}

Security and privacy protection are long-standing issues in wireless systems. Wireless signals propagate and are collected in open space, making them vulnerable to interference, spoofing, and malicious attacks. Minute perturbations applied to wireless signals may affect the quality of representations extracted and amplify errors in localization results. Raw channel measurements used for localization may contain user information, such as device identifiers. High-accuracy wireless localization also requires large numbers of channel observations and position labels, which contain sensitive private information such as precise user locations, activity trajectories, user identities, and behavioral patterns. During collection, transmission, and storage, these data may be misused, tampered, or leaked. Furthermore, locally adapted model parameters, which are easily overlooked, also belong to user privacy that should be protected. Federated learning, differential privacy, secure multi-party computation, and source data inaccessible adaptation approaches provide potential pathways for privacy protection, but they also introduce communication overhead, model performance degradation, and training stability issues. Therefore, how to protect raw channel data and location privacy while learning high-quality channel representations that support reliable localization is a key challenge for future wireless systems.

\section{Conclusion}
\label{sec:conclusion}

Learning-driven methods enable the extraction of location-related information from complex wireless observations, making high-accuracy localization feasible. Existing studies indicate that the performance and generalization of localization are not merely determined by increasingly complex network architectures or loss function designs. Instead, they depend more fundamentally on the quality and effectiveness of channel representations that encode location-related information from wireless observations. Based on this view, this survey models learning-driven high-accuracy wireless localization as a unified ``wireless observation--channel representation--location inference'' framework, and reviews related studies with channel representation as the organizing principle. Starting from channel observations, feature extraction, and representation learning, we evaluate the observation sources and representation construction methods used in wireless localization, and further organize existing techniques based on channel representations. We also examine the key challenges that must be resolved before these methods can be deployed in real systems.

\bibliographystyle{IEEEtran}
\bibliography{references} 

\end{document}